\documentclass[lettersize,journal]{IEEEtran}
\usepackage{amsmath,amsfonts}
\usepackage{algorithmic}
\usepackage{algorithm}
\usepackage{array}
\usepackage[caption=false,font=normalsize,labelfont=sf,textfont=sf]{subfig}
\usepackage{textcomp}
\usepackage{stfloats}
\usepackage{url}
\usepackage{balance}
\usepackage{verbatim}
\usepackage{graphicx}
\usepackage{comment}
\usepackage{cite}
\usepackage{xurl}
\usepackage[normalem]{ulem}

\usepackage{ifthen}
\usepackage{xcolor}
\usepackage{mathabx}
\newboolean{showcomments}
\setboolean{showcomments}{true}

\makeatletter
\newcommand{\mynote}[3]{%
  \ifthenelse{\boolean{showcomments}}{%
   \fbox{\bfseries\sffamily\scriptsize#1}%
   {\small$\blacktriangleright$\textsf{\emph{\color{#3}{#2}}}$\blacktriangleleft$}}%
  {%
   \@bsphack
   \@esphack
  }%
}
\makeatother

\definecolor{asparagus}{rgb}{0.53, 0.66, 0.42}
\newcommand{\pz}[1]{\mynote{Peipei}{#1}{blue}}
\newcommand{\zhuoping}[1]{\mynote{Zhuoping}{#1}{asparagus}}

\usepackage{xcolor}
\usepackage[hidelinks]{hyperref} 
\hypersetup{
  colorlinks=true,
  linkcolor=blue,
  urlcolor=blue,
  citecolor=magenta
}

\begin{document}

\title{\LARGE{Report for NSF Workshop on Algorithm-Hardware \\ Co-design for Medical Applications}
}

\author{
Peipei Zhou
,~Zheng Dong,~Insup Lee,~Aidong Zhang,~Robert Dick,~Majid Sarrafzadeh,\\
~Xiaodong Wu,~Weisong Shi,~Zhuoping Yang,~Jingtong Hu,~Yiyu Shi
\thanks{
Peipei Zhou and Zhuoping Yang are with Brown University.}
\thanks{
Zheng Dong is with Wayne State University. 
}
\thanks{
Insup Lee is with University of Pennsylvania
}
\thanks{
Aidong Zhang is with the University of Virginia
}
\thanks{
Robert Dick is with the University of Michigan.
}
\thanks{
Majid Sarrafzadeh is with UCLA.
}
\thanks{
Xiaodong Wu is with the University of Iowa
}
\thanks{
Weisong Shi is with the University of Delaware
}
\thanks{
Jingtong Hu is with the University of Pittsburgh
}
\thanks{
Yiyu Shi is with the University of Notre Dame
}
\thanks{
\IEEEauthorrefmark{1} Corresponding author: Peipei Zhou (peipei\_zhou@brown.edu)
}

}


\markboth{
}%
{
}


\maketitle

\begin{abstract} 
This report summarizes the discussions and recommendations from the NSF Workshop on Algorithm-Hardware Co-design for Medical Applications, held on September 26-27, 2024, in Pittsburgh, PA. The workshop assembled an interdisciplinary cohort of researchers, clinicians, and industry leaders to examine foundational challenges and develop a strategic roadmap for algorithm-hardware co-design in medical computing. 
The workshop focuses on four thematic areas: (1) teleoperations, telehealth, and surgical operations; (2) wearable and implantable medicine, including implantable living pharmacies; (3) home ICU, hospital systems, and elderly care; and (4) medical sensing, imaging, and reconstruction. 
This report calls for a fundamental shift in how next-generation medical technologies are conceived, designed, validated, and translated into practice. 
The report recommends that NSF sustain investment in shared standardized data infrastructures and compute infrastructures, 
develop clinic workflow-aware systems and human-AI collaboration frameworks, 
promote scalable validation ecosystems grounded in objective, continuous measures, and 
physics-informed
evaluation, 
and enable safe, accountable, and resilient 
platforms, including virtual-physical healthcare ecosystems, to de-risk translational pathways.
The workshop information can be found on the website:\{{\url{https://sites.google.com/view/nsfworkshop}}\}. 
\end{abstract}


\section{Executive Summary}

The intersection of advanced computing and clinical medicine has reached a critical inflection point. While breakthroughs in artificial intelligence and sensing have created immense potential for personalized care, the translation of these technologies into real-world clinical impact is currently obstructed by 
insufficient frameworks for robust, real-time, and human-centered medical AI, compounded by data variability and extreme patient variability, 
fragmented care ecosystems, 
and a widening ``valley of death'' between academic prototypes and commercially deployable solutions.


This report summarizes the discussions and recommendations from the NSF Workshop on Algorithm-Hardware Co-design for Medical Applications, held on September 26-27, 2024, in Pittsburgh, PA.
The workshop focuses on four thematic areas: (1) teleoperations, telehealth, and surgical operations; (2) wearable and implantable medicine, including implantable living pharmacies; (3) home ICU, hospital systems, and elderly care; and (4) medical sensing, imaging, and reconstruction.
Within each thematic chapter, the workshop discussions are organized into five components: (i) overview and motivation; (ii) representative existing efforts and promising directions; (iii) key challenges and showstoppers; (iv) cross-cutting themes and research gaps; (v) future directions and recommendations to NSF. 
This structure ensures a comprehensive examination of both current capabilities and fundamental barriers, connecting state-of-the-art advances to unresolved technical, translational, validation, and system-level challenges. 
In addition to the recommendations articulated within each research theme, Chapter~\ref{sec:general} synthesizes cross-cutting insights across all four thematic areas to establish a cohesive, system-level framework for NSF investment focused on resilience, scalability, safety, and long-term clinical impact.

This workshop brought together health practitioners, medical application domain experts, data analysts, algorithm developers, computer architects, and hardware designers to architect a new paradigm: holistic algorithm-hardware co-design that treats medical systems as unified cyber-physical entities. 
By synchronizing the design of sensors, specialized accelerators, human-variability-aware algorithms, and human-in-the-loop systems, the research community can move beyond episodic, hospital-centric care toward a future of continuous, proactive, and resilient health management.

\noindent\uline{The workshop explored four key themes:}

\begin{enumerate}
    \item Teleoperations, telehealth, and surgical operations (Chapter~\ref{sec:teleoperation}). 
    Focusing on the ``safety-latency gap,'' this theme emphasizes the need for tightly integrated sensing, computation, control, and networking systems that are seamlessly aligned with clinical workflows to satisfy the safety, reliability, and real-time constraints of remote surgery and telementoring. 
    \item Wearable and implantable medicine, including implantable living pharmacies (Chapter~\ref{sec:wearable}). Shifting the focus from ``fitness tracking'' to living pharmacies, this theme prioritizes longevity-first design and sub-milliwatt systems-on-chip (SoCs) that tightly integrate sensing, on-chip intelligence, and communication to enable safe, adaptive, and closed-loop operation under extreme in-body power, size, and thermal constraints.
    \item Home ICU, hospital systems, and elderly care (Chapter~\ref{sec:homeICU}). Transitioning critical care into the residential living room, this theme identifies the home as a first-class engineering constraint, requiring 
    objective, continuous measures that replace subjective self-reports with scalable ground truth and advance from detection alone to validated, closed-loop mental and social health interventions.
    \item Medical sensing, imaging, and reconstruction (Chapter~\ref{sec:medicalSensing}). 
    Addressing the exponential growth of four-dimensional data, this theme advances 
    a two-fold strategy that unifies algorithmic and system-level innovation. 
    On the algorithmic front, it emphasizes annotation-efficient learning, foundation models, diffusion-based segmentation, and expert-in-the-loop quality assurance to ensure accuracy, robustness, and clinical trust; on the system level, it promotes modular, software-defined imaging platforms and virtual-physical ``medical metaverses'' that enable large-scale validation and accelerate safe clinical translation.
\end{enumerate}

\noindent\uline{Core Recommendations to the NSF.} The workshop synthesized these themes into a strategic roadmap for federal investment, focused on system-level resilience and long-term adaptation:
\begin{enumerate}

\item \textbf{Invest in ``Human-Centered Resilience''.} Rather than designing devices that operate only under ideal conditions, medical technologies should be tough by design. This entails ensuring reliable operation despite network outages, sensor drift, or partial hardware failures. In addition, such systems should be evaluated in ``living labs'' that reflect the physiological variability and environmental complexity of the human body.

\item \textbf{Clear the Path from Lab to Hospital.} Many promising innovations fail to translate due to prohibitive manufacturing costs or regulatory barriers. Addressing these real-world constraints early in the research lifecycle is essential. Low-barrier design challenges can enable rapid, safe prototyping and evaluation, allowing researchers to assess feasibility before committing to long and costly clinical trials.

\item \textbf{Make Medical Hardware Plug-and-Play.} The prevailing one-size-fits-all paradigm limits upgradability and longevity. Modular, reconfigurable medical systems, where components can be easily swapped or upgraded, offer a more sustainable alternative. Such architectures reduce maintenance costs, simplify repairs, and mitigate technological obsolescence as new capabilities emerge. 

\item \textbf{Use Data That Reflects Real Life.} Clinical decision-making often relies on self-reported surveys that can be subjective or incomplete. Investment is needed in digital biomarkers that provide objective, continuous measurements of mental and physical health. Equally important is the collection of longitudinal data capturing real-world behaviors and social determinants that influence health outcomes over time.

\item \textbf{Connect the Body and the Mind through Physical AI.} Health extends beyond physiological signals to include emotional state and lived experience. 
We must support the development of Physical AI-driven ``living pharmacies'', 
adaptive therapeutic systems
that use closed-loop reasoning to modulate therapy in response to biological, emotional, or social cues. 
By embedding intelligence directly 
at the device-tissue or device-body interface,
these systems can perform real-time sensor recalibration and autonomous adjustment, integrating the biological system and the individual as a functional whole. 


\item \textbf{Create AI That Works Everywhere.} 
Medical AI spans a spectrum of computational scales, from high-performance systems that support complex tasks such as MRI interpretation to ultra-efficient models designed for resource-constrained wearable and implantable devices
Flexible, multimodal AI architectures can ensure consistent quality of care across settings, from advanced hospitals to rural and home-based environments.

\item \textbf{Protect Privacy While Sharing Progress.} Future medical systems should be designed to share insights rather than raw data. This enables clinicians to receive actionable information without exposing sensitive personal data. Achieving this vision requires equipping academic researchers with computing infrastructure comparable to that of industry, empowering them to lead advances in privacy-preserving, open medical research.

\end{enumerate}
\section{Background
}


Medical applications serve as an integral component of healthcare and they are the nexus of rigorous academic research and day-to-day clinical practice, reflecting a commitment to improving human health on both an individual and societal level. 
When considering the realm of medical applications, the importance of electronic design automation (EDA) research cannot be overstated. 
EDA provides the means to attain unprecedented precision in algorithm and hardware development, enhancing functionality and ensuring stringent adherence to safety standards. The end products of this process are meticulously crafted instruments capable of facilitating life-saving interventions and augmenting patient quality of life. 
Personalized medicine represents another significant frontier in which medical applications display their importance. 
Here, EDA enables the tailoring of medical devices, prosthetics, and treatment plans to the individual patient’s unique requirements, promoting optimal therapeutic outcomes. 

\begin{figure}
    \centering
    \includegraphics[width=1\linewidth]{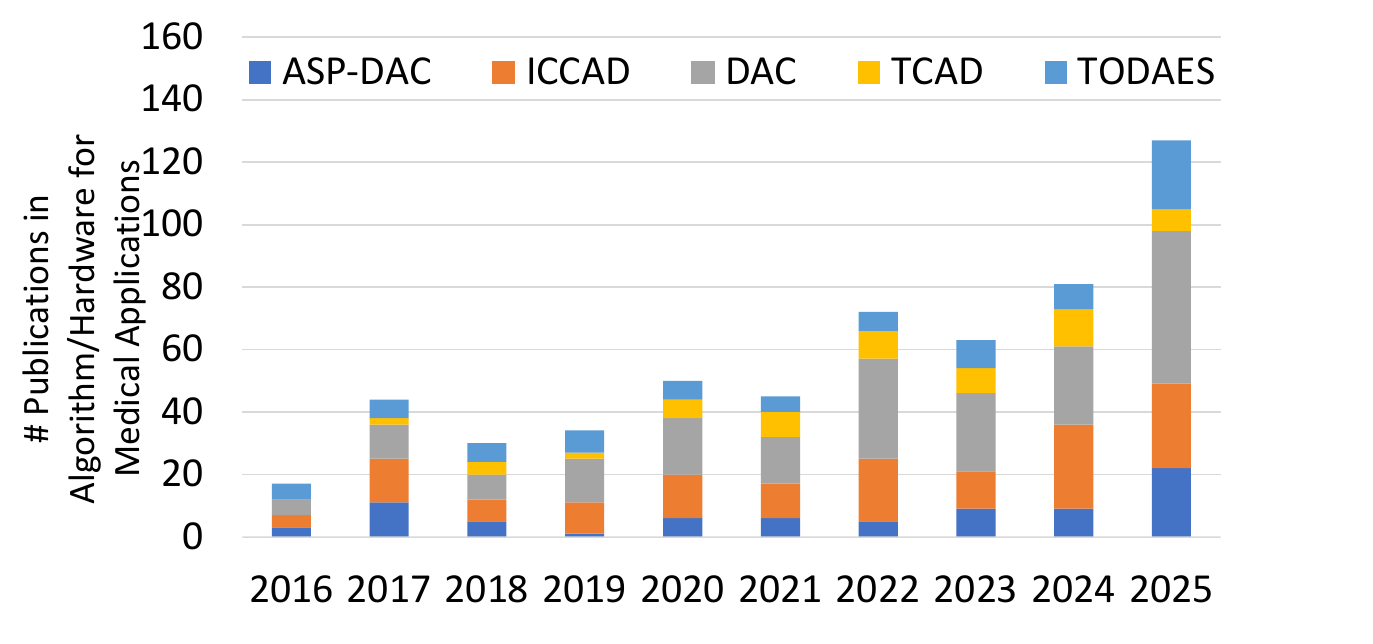}
    \caption{
    Growth in publications related to algorithm-hardware co-design for medical applications across leading EDA conferences and journals from 2016 to 2025.
    }
    \label{fig:motivation_example}
    \vspace{-10pt}
\end{figure}

There has been increasing interest and growing demand for algorithm-hardware co-design for medical applications in EDA. 
This trend is reflected in the rising number of related publications in major EDA conferences and journals, as well as in rapid innovation within the medical devices industry and research community.
Figure~\ref{fig:motivation_example} illustrates the number of publications on algorithm-hardware co-design for medical applications across major EDA journals and venues, including TODAES, TCAD, DAC, ICCAD, and ASPDAC, from 2016 to 2025. 
The figure shows a steady and pronounced increase in publications over this period, with particularly strong growth after 2019. 
Contributions appear across both conferences and journals, indicating broad and sustained interest from the EDA research community. 
This upward trend reflects the growing importance of medical applications as a key driver for algorithm-hardware co-design research.
From an industry perspective, according to Global Market Insight, the medical application is the top 1 increasing application from 2017 to 2024 in the Global EDA market~\cite{GMI}. 
At the same time, advances in artificial intelligence (AI) and semiconductor technology have led to pioneering medical device designs that can help diagnose diseases, assist those with disabilities, and provide real-time patient care for all age groups.
In light of these trends, the NSF Workshop on Algorithm-Hardware Co-design for Medical Applications was organized to bring together experts from academia, industry, and clinical practice.

The workshop included 15 invited speakers, four discussion panels with 21 panelists, and four roundtable discussions, covered four themes, and attracted more than 110 attendees.

\section{Research Theme 1: Teleoperations, Telehealth, and Surgical Operations
}
\label{sec:teleoperation}

\subsection{Overview and Motivation}

Teleoperations, telehealth, and surgical robotic systems are no longer futuristic concepts; they are essential responses to the modern healthcare crisis of workforce shortages and the ``specialist gap'' in rural or underserved areas. While advances in sensing and communication have brought us to the cusp of a remote-care revolution, the transition from isolated technological success to system-wide clinical impact remains the central challenge of the field.

\underline{The Co-Design Imperative.} As evidenced by representative efforts, ranging from high-speed retinal imaging at the primary care edge to concentric tube robots navigating needle-sized anatomical paths, the most successful systems are those that employ a holistic algorithm--hardware co-design. These systems demonstrate that medical-grade performance cannot be achieved by software alone; it requires specialized sensing pipelines that manage latency at the hardware level and Compute-in-Memory (CiM) architectures that allow sophisticated AI to run on energy-constrained surgical tools.

\underline{Beyond the Laboratory: Showstoppers and Gaps.} Despite these technical milestones, the ``Valley of Death'' between academic prototyping and clinical reality is wider than in most other computing fields. The workshop identified several critical \emph{showstoppers} that the research community must address:

\begin{itemize}
    \item \textbf{Workflow is critical:} Algorithmic accuracy is insufficient if the system increases clinician cognitive load or disrupts established medical protocols.
    
    \item \textbf{The ``Human'' Variable:} Existing design frameworks often fail to account for the extreme physiological diversity of patients or the nuanced, adaptive behavior of expert surgeons.
    
    \item \textbf{Data Mismatch:} While resources like the UK Biobank and AI-READI provide a foundation, there is a persistent gap in ``research-ready'' datasets that capture the messy, real-time failures of edge-based medical devices.
\end{itemize}

\underline{A Call for System-Level Research (NSF recommendations).} Moving forward, the focus must shift from device-level innovation to system-level resilience and translation. This requires a fundamental pivot toward:

\begin{itemize}
    \item \textbf{Human-in-the-loop Modeling:} Developing formal abstractions that treat the clinician and patient as dynamic components of a closed-loop control system.
    
    \item \textbf{Infrastructure for ``Pre-Clinical'' Innovation:} Creating pathways for hardware and computing researchers to \emph{de-risk} medical technologies, addressing manufacturability, security, and regulatory hurdles, long before they reach expensive clinical trials.
    
    \item \textbf{Cross-Agency Synergy:} Leveraging NSF’s expertise in fundamental engineering and translation (via the TIP Directorate) to transform existing NIH-funded data into actionable, co-designed architectures.
\end{itemize}

The following sections detail the landscape of these emerging technologies, the cross-cutting gaps that hinder their progress, and a strategic framework for future investment to ensure that the next generation of medical systems is as deployable as it is innovative.

\subsection{Representative Existing Efforts and Promising Directions}

This section summarizes representative existing efforts discussed in the workshop that illustrate current approaches and system-level advances in teleoperations and telehealth.


One representative existing effort discussed in the workshop focused on enabling diabetic retinopathy screening in primary care settings through {on-device intelligence} implemented on a retinal camera. 
The system uses high-speed, high-sensitivity image sensors (120 fps) tightly integrated with the system-on-chip to perform burst-image merging and real-time image quality assessment, tightly integrating sensing, computation, and the software pipeline.
Diabetes mellitus affects more than 40 million~\cite{CDC_DM_Stats} individuals in the United States and imposes an estimated annual cost of approximately \$412 billion~\cite{ ADA_Economic_Burden} due to direct medical expenses and lost productivity, while diabetic retinopathy (DR) impacts approximately 35\% of patients and remains a leading cause of preventable blindness \cite{DR_Prevalence}.
Diabetic retinopathy remains a leading cause of preventable blindness, yet screening rates remain low due to reliance on specialty ophthalmology clinics, high equipment costs, and the need for trained operators. 
These barriers are particularly acute in small and medium-sized practices that lack dedicated imaging infrastructure.
The presented system, as described in recent work~\cite{Kavusi2025Retina}, consists of a nonmydriatic retinal camera that performs substantial processing directly on the device and interfaces with cloud-based software for downstream workflows. 
On the camera, high-speed image sensors capture a burst of retinal images within a single flash. 
These images are processed locally using proprietary burst-imaging and image-compilation algorithms that merge multiple frames into a single high-fidelity retinal image, improving robustness to cataracts, small pupil size, eye motion, and other common artifacts in diabetic populations. 
In addition, the camera executes an on-device image quality control algorithm that evaluates retinal visibility in real time and determines whether image recapture is needed, providing immediate feedback during acquisition.
After on-device image compilation and quality assessment, the resulting retinal images are transmitted to a cloud-based software platform 
that accepts, stores, and routes digitized retinal images for grading and supports integration with electronic health record (EHR) systems.
This separation of responsibilities enables time-critical sensing and quality control to occur at the edge, while leveraging the cloud for scalability and interoperability with existing health information systems. 
By explicitly partitioning computation between the camera and the cloud, this effort demonstrates how algorithm-hardware co-design can reduce operator dependence, improve image quality at capture time, and support scalable telehealth screening in primary care environments.

Another representative existing effort discussed in the workshop focused on \textbf{surgical robotic systems}~\cite{2025roboticSurgeryReview} designed to enable minimally invasive procedures in anatomically constrained environments. 
This line of work centers on continuum robotic systems~\cite{websterPaper1, websterPaper2, websterPaper3,websterPaper4,websterPaper5,websterPaper6}, particularly concentric tube robots that form needle-sized robotic manipulators, which provide enhanced dexterity and reach beyond conventional rigid surgical instruments by operating through natural orifices or small access points. 
A representative academic example is the development of concentric tube robotic systems~\cite{websterPaper1} for performing suturing during radical prostatectomy, in which multiple precurved elastic tubes are deployed through an endoscope to form a steerable robotic manipulator capable of precise needle positioning and suturing within confined anatomical spaces. Experimental studies in biological tissue demonstrate the feasibility of performing transurethral anastomosis using this robotic approach, highlighting its potential to reduce surgical complexity and improve access compared to conventional surgical systems. Beyond academic prototypes, this effort also highlighted the translation of concentric tube robotic technology into clinical systems through commercialization. The Virtuoso Surgical Robotic System~\cite{newsVirtuoso}, developed by Virtuoso Surgical Inc.~\cite{Virtuoso}, 
is founded on concentric tube robotics originally conceived in this research program and has recently been used in first-in-human procedures for endoscopic en bloc bladder tumor resection~\cite{newsVirtuoso}. 
These early clinical cases demonstrate how algorithm-hardware innovations in surgical robotic systems can progress from laboratory research to real-world clinical impact through system-level integration, clinical collaboration, 
and regulatory pathways, 
underscoring the importance of coordinated co-design across algorithms, hardware, control, and clinical context for successful translation into practice.

In addition to deployed and translational systems, workshop discussions also highlighted promising directions in \textbf{autonomous intervention}, \textbf{teleoperated surgical robots}, and \textbf{computer-assisted robotic surgery (CARS)}~\cite{2025roboticSurgeryReview}. 
Recent work on autonomous steerable needle systems~\cite{websterPaper7} demonstrates the feasibility of navigating needles through living tissue while accounting for anatomical obstacles and physiological motion, pointing to future opportunities for algorithm-hardware co-design in computer-assisted robotic surgery.

A third representative effort focuses on \textit{magnetically actuated medical robotics}~\cite{Zemmar1,Zemmar2,Zemmar3} for minimally invasive neurosurgery, addressing fundamental limitations of conventional rigid, straight-line instruments used in procedures such as deep brain stimulation (DBS). 
Instead of relying on mechanical pushing forces that can damage tissue and restrict trajectories, this approach uses externally generated magnetic fields to wirelessly steer flexible, magnetically tipped needles along smooth, curved paths through soft brain tissue. 
By decoupling steering forces from direct tissue interaction, the system reduces targeting errors caused by brain shift, tissue heterogeneity, and anatomical variability, while enabling precise access to deep or hard-to-reach regions. 
Proof-of-concept studies demonstrate that magnetically guided needles can follow complex trajectories and reach multiple targets from a single entry point under human-in-the-loop control, minimizing repeated insertions and associated trauma.

A fourth representative effort explored the enablement of human-AI collaborative surgical telementoring, integrating real-time medical imaging with interactive AI assistance under stringent system constraints. This research highlights the critical need for visual guidance in high-stakes cardiac interventions, such as aortic valve replacement, intracardiac catheter navigation, and cardiac mapping, where anatomical complexity, physiological motion, and safety-critical demands pose significant barriers. To address these challenges, the presented systems utilize real-time cardiac cine magnetic resonance imaging (MRI) guidance~\cite{chen2023ame, lu2022rt, dong2020deu, dong2022deu, wang2019msu} and contrast-enhanced ultrasound for myocardial segmentation and automatic perfusion assessment~\cite{li2023novel, chen2022myocardial, zeng2021hardware}. These examples illustrate a necessary paradigm shift: imaging, learning algorithms, and hardware must be co-designed to satisfy rigorous latency and accuracy requirements. Notably, this body of work includes some of the earliest demonstrations of human-AI collaborative telementoring for congenital heart disease~\cite{ShiPaper1}, showcasing how AI can augment expert guidance during live surgical procedures. Beyond procedural assistance, the workshop addressed the broader challenge of scaling AI for personalized healthcare. This includes emerging techniques for on-device Large Language Model (LLM) fine-tuning~\cite{qin2025tiny, qin2024enabling} and the enhancement of federated learning through personalized optimization~\cite{jia2024personalized, xu2025fair}. As AI/ML model sizes and computational complexities grow exponentially~\cite{jia2023importance}, the workshop underscored Compute-in-Memory (CiM) as a transformative direction for resource-constrained systems. Recent advances~\cite{yan2023improving, qin2025nvcim, qin2024tsb, yan2024compute, qin2024special, yan2024u} demonstrate that CiM architectures can significantly reduce hardware resource demands and energy consumption, positioning them as a cornerstone for future real-time medical diagnosis and edge-deployed clinical assistance.

A fifth representative effort discussed in the workshop
focuses on using AI to extract high-level, clinically meaningful information from complex and often unstructured biomedical data. This research addresses a critical gap in current systems: the inability to systematically incorporate Social Determinants of Health (SDoH) into clinical decision-making. Representative work in this area includes the use of knowledge-augmented LLMs to boost the extraction of SDoH from electronic health records~\cite{gong2024context, gong2025boosting} and the integration of this information to improve the diagnosis of Alzheimer’s disease~\cite{noshin2025integrating}. Furthermore, this line of research demonstrates how uncovering latent patterns in SDoH can enhance risk prediction for alcohol use disorder~\cite{meyer2025enhancing} and provide deeper insights into high-stakes scenarios such as veteran suicide risk~\cite{chen2025leveraging}. In addition to data extraction, this effort addresses growing concerns surrounding the safe and reliable adoption of LLMs in medical systems. As these models move toward clinical use, verifying their outputs becomes a primary safety requirement. Recent contributions have examined whether LLMs can generate verifiable medical content and provide trustworthy clinical reasoning~\cite{wang2025medcite}, alongside rigorous evaluations of their reliability and safety risks in real-world settings~\cite{yang2024ensuring}. To support these evaluations, the work introduces specialized benchmarks, such as MedCalc~\cite{khandekar2024medcalc}, designed to assess the accuracy of LLMs in complex medical calculation tasks. Collectively, these efforts underscore the necessity of building trustworthy AI layers that can translate messy, real-world data into verified, context-aware clinical insights.

\subsection{Key Challenges and Showstoppers}

Discussions during panels and roundtable sessions identified several key challenges and showstoppers that continue to limit the translation of telehealth and teleoperation systems into real-world clinical practice.

\begin{itemize}

  \item \textbf{Mismatch between available clinical data and algorithm needs.}
  Much of the data captured in electronic health records is optimized for billing and compliance rather than diagnosis or workflow analysis, limiting its usefulness for training and validating robust clinical AI systems.

  \item \textbf{Real-time, safety-critical constraints in AI-assisted medical systems.}
  Across examples, discussions identified practical challenges that limit deployment, including the need for real-time response under constrained hardware resources, privacy and latency requirements that motivate edge computation, noisy and variable input data, large inter-patient variability, limited tolerance for performance-accuracy trade-offs in safety-critical settings, and the difficulty of effective human-model collaboration during live procedures.

  \item \textbf{Security and trust across the full system stack.}
  Panelists identified security as a major adoption barrier, extending beyond software to include communication links, hardware, and even mechanical components. 
  Concerns included system backdoors, unintended functionality, and patient safety risks in teleoperated settings.

  \item \textbf{Extreme human variability and personalization requirements.}
  Human variability, both across patients and over time for the same individual, was repeatedly identified as a challenge, particularly for implantable and chronic-use systems that must operate reliably for decades without intervention or charging.

  \item \textbf{Workflow integration as a dominant adoption barrier.}
  Participants emphasized that algorithmic accuracy alone is insufficient for deployment. Systems that disrupt existing clinical workflows or increase provider workload face slow adoption, regardless of technical performance.

  \item \textbf{Device-level success does not translate into system-level clinical impact.}
  Discussions emphasized that a technically sound product or high algorithmic accuracy alone is insufficient to achieve meaningful clinical impact. Participants highlighted that outcomes depend on system-level design, integration, and evaluation. For example, providing immediate feedback within clinical workflows led to approximately a threefold increase in post-intervention follow-up, while adapting retinal imaging technology for individuals with albinism resulted in orders-of-magnitude cost reduction and significantly higher imaging success rates. At the same time, participants noted a lack of systematic approaches for modeling and evaluating clinical impact at the system level. Many existing efforts focus on device-level performance or algorithmic accuracy, while downstream effects such as follow-up behavior, care pathways, cost structures, and accessibility are rarely treated as first-class design objectives. As a result, it remains difficult to predict, measure, or compare real-world impact across telehealth solutions, even when individual components perform well.

  \item \textbf{Regulatory and IRB timelines as a fundamental barrier.}
  Participants emphasized that teleoperation and remote surgery face unusually long institutional review board (IRB) timelines, often requiring years before approval, which significantly slows iteration, validation, and deployment compared to other medical technologies.

  \item \textbf{Misalignment between academic prototypes and commercial translation requirements.}
  Workshop discussions highlighted that early-stage surgical robotic technologies may initially attract interest from large medical device companies, but such interest often declines when systems remain at the laboratory prototype stage. Panelists emphasized that academic teams frequently under consider financial value propositions, manufacturability, regulatory strategy, and realistic cost and timeline milestones required to transition from a prototype to a commercial product. As a result, promising technologies can struggle to move beyond proof-of-concept despite strong technical merit.

\end{itemize}

\subsection{Cross-Cutting Themes and Research Gaps}

Across invited talks, panel discussions, and roundtable sessions, several cross-cutting themes emerged that highlight both progress and remaining gaps in algorithm-hardware co-design for telehealth and teleoperation systems. 
In particular, discussions revealed recurring patterns related to data availability, system integration, and evaluation practices that extend beyond any single application or deployment setting.

\begin{itemize}

\item \textbf{High-quality, research-ready medical datasets remain scarce.}
Discussions noted that only a limited number of large-scale, well-curated medical datasets are broadly available to the research community. 
A prominent example is \textit{UK Biobank}~\cite{Bycroft2018UKBiobank}, a prospective cohort study with deep genetic and phenotypic data collected from approximately 500,000 participants across the United Kingdom. The resource includes rich multimodal information such as biological measurements, lifestyle indicators, biomarkers, large-scale imaging, longitudinal health records, and genome-wide genotype data, enabling population-scale analysis of complex traits and disease associations. 
More examples include \textit{AI-READI}~\cite{AI-READI}, one of four Data Generation Projects funded by the National Institutes of Health (NIH) Common Fund program Bridge2AI~\cite{bridge2ai}, which aims to publicly release multimodal, AI-ready data for studying the pathogenesis and salutogenesis of type 2 diabetes mellitus,
and MIMIC-III~\cite{johnson2016mimic}, a widely used, de-identified critical care database containing clinical, physiological, and treatment records from ICU patients.
The small number of such comprehensive efforts highlights a broader scarcity of accessible, high-quality datasets that support algorithm-hardware co-design and system-level evaluation in medical applications.

\item \textbf{High-quality medical data requires more than raw clinical records.}
Discussions emphasized that most clinical data collected in electronic health record (EHR) systems is optimized for billing and compliance rather than algorithm development or system evaluation. 
As a result, diagnostic detail is often limited, workflow-related information is sparse, and failure cases are not systematically captured. These limitations reduce the usefulness of raw clinical data for designing, validating, and deploying end-to-end telehealth systems, even when large volumes of data are available.

\item \textbf{Limited shared data infrastructure and standardization across institutions.}
While many institutions collect large volumes of medical data, differences in data formats, modalities, and metadata severely limit data sharing and reuse, hindering community-wide progress and reproducibility.

\item \textbf{Insufficient frameworks for robust, real-time, and human-centered medical AI.}
Workshop discussions revealed gaps in methods that jointly address hardware constraints, data variability, and human interaction in real-time medical systems. 
Existing approaches often lack principled ways to manage inter-patient variability, reason about performance-accuracy trade-offs under safety constraints, and support reliable human-AI collaboration, particularly when computation must occur at the edge due to privacy and latency requirements.


\item \textbf{Limited consideration of clinical workflows in algorithm-hardware co-design.}
Across existing efforts discussed in the workshop, clinical workflows are often treated as external constraints rather than first-class design objectives. Algorithm and hardware development frequently prioritizes accuracy, performance, or efficiency in isolation, with limited integration of workflow modeling, provider interaction, and operational constraints. As a result, many technically strong systems fail to align with real-world clinical practice, contributing to persistent adoption challenges observed across telehealth and teleoperation applications.

\item \textbf{Lack of models and tools for human-aware and human-in-the-loop system design.}
Discussions revealed a gap in formal models and design tools that explicitly account for both human variability and active human participation in safety-critical medical systems. 
Existing approaches often lack abstractions to represent patient-to-patient differences, clinician variability and expertise, physiological and anatomical diversity, and long-term changes over time, limiting robustness and personalization across populations. 
In parallel, there is a lack of principled frameworks for human-in-the-loop operation, including shared control between clinicians and AI systems, intervention and override mechanisms, uncertainty communication, and assignment of responsibility during teleoperation and surgical robotics. The absence of unified models that capture both human variability and real-time human-system interaction constrains the ability to design, test, and evaluate teleoperated and robotic medical systems prior to clinical deployment.

\item \textbf{Insufficient testing, debugging, and verification methods for safety-critical medical systems.}
Participants noted the lack of systematic approaches for testing, debugging, and verifying algorithm-hardware co-designed systems, particularly when outputs directly influence human health and safety.

\item \textbf{Limited integration of commercialization and regulatory considerations into early-stage research.}
Across invited talks, panel discussions, and roundtable sessions, a recurring theme was the lack of research frameworks that incorporate manufacturing constraints, regulatory requirements, and economic feasibility into algorithm-hardware co-design. 
These factors are often treated as downstream concerns, even though they strongly influence whether technologies can progress from academic prototypes to clinically validated products. This gap limits the ability to assess translation readiness and slows the pathway from research innovation to real-world clinical impact.

\end{itemize}

\subsection{Future Directions and Recommendations to NSF}

While NIH plays a primary role in supporting clinical studies and data generation, workshop discussions highlighted several complementary opportunities for NSF to advance algorithm-hardware co-design research by focusing on methods, abstractions, and system-level design frameworks that build on these clinical resources.

\begin{itemize}

\item \textbf{Expand the availability of research-ready medical datasets for algorithm-hardware co-design.}
Discussions highlighted that only a limited number of large-scale, well-curated medical datasets are broadly available to the research community, with efforts such as UK Biobank and AI-READI serving as representative examples. NSF could support research activities that improve the accessibility, usability, and sustainability of such datasets for algorithm-hardware co-design, including standardized interfaces, documentation, and benchmarks that facilitate system-level evaluation.

\item \textbf{Enable algorithm-hardware co-design research that builds on existing NIH-funded data resources.}
While large-scale clinical data generation is primarily supported by NIH, workshop discussions identified a clear role for NSF in advancing algorithm-hardware co-design research that leverages these datasets. In particular, NSF could support methods, tools, and abstractions that transform existing clinical data into forms usable for system-level design, such as machine-interpretable annotations, failure characterization, and longitudinal behavior modeling, without duplicating clinical data collection efforts.

\item \textbf{Enable shared data infrastructure and standardization without requiring clinical trials.
}
NSF could lower barriers by supporting data standardization, software infrastructure, and community platforms that enable sharing and reuse of medical data, even in the absence of immediate clinical evaluation, with pathways to later NIH or FDA-supported studies.

\item \textbf{Support research on secure, fault-tolerant, and resilient medical systems.}
NSF could encourage foundational research that addresses security, fault tolerance, graceful degradation, and long-term reliability from the ground up, spanning hardware, software, communication, and mechanical components.

\item \textbf{Advance workflow-aware and system-level design and evaluation methodologies.}
Beyond data availability, participants emphasized the need for research methodologies that explicitly incorporate clinical workflows, provider interaction, care pathways, and system constraints into the design and evaluation. 
NSF could promote approaches that treat workflows and system constraints (latency, privacy, hardware limits, and safety requirements) as first-class design objectives and develop evaluation frameworks that link algorithm and hardware design choices to system-level outcomes, enabling more effective translation of telehealth and teleoperation systems into practice.

\item \textbf{Invest in human-aware and human-in-the-loop system models.}
Future NSF efforts could support research on digital twins, human-aware modeling, and adaptive systems that explicitly account for 
patient-to-patient differences, clinician variability and expertise, physiological and anatomical diversity, and evolving user behavior over time.
Also, NSF could encourage the development of human-in-the-loop frameworks that support shared control between clinicians and AI systems, intervention and override mechanisms, uncertainty communication, and responsibility allocation in safety-critical settings. 

\item \textbf{Create early-stage, impact-oriented entry mechanisms for medical innovation.
}
Workshop discussions noted that NIH provides tiered funding mechanisms (e.g., R21, R01) that support progressively mature biomedical research, while early exploratory opportunities remain limited for computing and hardware researchers seeking to test high-risk medical ideas. 
In addition, NSF could sponsor design contests, focused workshops, and collaborative challenge programs that directly connect engineering researchers with medical professionals.
NSF could establish small-scale, low-barrier programs or design challenges that enable interdisciplinary teams of hardware, algorithm, and medical professionals to prototype and assess feasibility and potential impact without requiring immediate clinical evaluation. 
These efforts could emphasize rapid iteration, measurable system-level impact, and early engagement with clinicians, while encouraging technology-industry and hospital partnerships and leveraging cost-sharing or matching support from medical centers and healthcare facilities. Clear handoff pathways could be defined in which promising ideas transition to other NSF programs or to NIH-supported clinical evaluation, allowing investigators to de-risk concepts at the pre-clinical stage before entering more resource-intensive studies.

\item \textbf{Strengthen mid-stage research-to-translation pathways for medical technologies.}
Workshop discussions highlighted that many promising medical systems stall between initial prototypes and clinical or commercial deployment due to gaps in manufacturability planning, regulatory readiness, economic feasibility, and milestone-driven translation strategies. 
NSF is well positioned to address this challenge by building on its existing translation-focused efforts, including programs within the Technology, Innovation and Partnerships (TIP) Directorate such as Accelerating Research Translation (ART), I-Corps, and the NSF Translation to Practice (TTP) initiative. 
Continued and expanded support for translation-focused research activities that connect technical innovation with regulatory, manufacturing, and commercialization considerations would help reduce risk, improve transition readiness, and accelerate the pathway from research outcomes to real-world medical impact.

\end{itemize}

\section{
Research Theme 2: Wearable and Implantable Medicine, Including Implantable Living Pharmacies
}
\label{sec:wearable}



{\subsection{Overview and Motivation}}

The landscape of personal healthcare is shifting from episodic, hospital-centric interventions toward a model of continuous, proactive, and personalized management. Central to this transition are wearable and implantable medical systems that act as persistent digital guardians, bridging the gap between clinical visits and daily life. While early-generation wearables focused primarily on consumer-grade fitness tracking, the next frontier, implantable living pharmacies and high-fidelity neural interfaces, demands a fundamental rethink of how algorithms and hardware are co-designed for the extreme environment of the human body.

\underline{The Challenge of In-Body Integration.} Unlike standard mobile computing, wearable and implantable devices operate under a triple constraint of near-zero power consumption, microscopic form factors, and absolute thermal safety. As demonstrated by representative efforts discussed in Sec.~\ref{sec:representive effort} such as contactless vital sign monitoring using 3D cameras and sub-milliwatt neural SoCs for prosthetic control, the hardware must do more than simply host the algorithm. Instead, it must intelligently filter, compress, and prioritize massive streams of physiological data to prevent data drowning and premature battery exhaustion. 
Complementary to such wearable platforms,
implantable bidirectional brain interfaces introduce even stricter constraints, requiring simultaneous sensing and stimulation (e.g., optogenetic brain modulation) without mutual interference.

\underline{From Static Devices to Adaptive Digital Twins.} A key insight from the workshop discussion is that medical devices can no longer be fixed at the time of manufacture. Because the human body is inherently dynamic, e.g., aging, healing, and adapting over time, both hardware and software must support long-term programmability and in-situ personalization. The emergence of Cardiovascular Digital Twins~\cite{jafariPaper1} provides a unifying vision for this paradigm, in which real-time sensor data feeds hierarchical, physics-informed models capable of predicting disease progression before clinical symptoms manifest.

\underline{Addressing the ``Longevity Gap'' (NSF recommendations).} Translating these innovations into deployable, life-saving systems requires overcoming several critical showstoppers, most notably the lack of realistic in-body testing models and the difficulty of upgrading algorithms in implanted hardware without invasive surgical procedures. To address these challenges, the workshop identified a critical role for NSF in advancing the following research directions:

\begin{itemize}
    \item \textbf{Longevity-First Design:} Shifting the design focus toward systems that can safely retrain, adapt, and evolve over decades-long deployment horizons.
    \item \textbf{Privacy-Preserving Edge Intelligence:} Developing ``transmit-information-not-data'' architectures that deliver actionable clinical insights while strictly adhering to HIPAA and related regulatory constraints.
    \item \textbf{Bio-Realistic Validation:} Moving beyond simplified electrical test circuits to create validation environments that capture the full complexity of tissue motion, physiological variability, and long-term biological interaction.
\end{itemize}

By fostering a research ecosystem that prioritizes graceful degradation, usability, and lifelong adaptation, the community can move beyond passive monitoring and toward true living pharmacies, i.e., autonomous, closed-loop systems capable of detecting and treating illness from within the human body.

\subsection{Representative Existing Efforts and Promising Directions}\label{sec:representive effort}

This section summarizes representative existing efforts discussed in the workshop that illustrate current approaches and system-level advances in wearable and implantable medicine, including implantable living pharmacies.


One representative effort discussed in the workshop focused on \textit{contactless vital sign monitoring (CVSM)} using \textbf{3D camera-based sensing systems}.
This work leverages depth cameras embedded in smartphones and tablets to measure physiological signals such as heart rate, respiratory rate, blood oxygen saturation, and body motion without requiring physical contact or wearable sensors. 
By building on widely available consumer 3D imaging hardware, the approach aims to enable scalable and cost-effective monitoring across hospital, intermediate care, home, and elderly care settings.
The presented system integrates near-infrared imaging, depth sensing, and signal processing to address key challenges in remote physiological measurement. 
Specifically, depth information from 3D cameras is used to compensate for out-of-plane motion, while active near-infrared illumination reduces sensitivity to ambient lighting variations and melanin-related absorption differences across skin tones. 
Building on these sensing modalities, the system combines facial blood flow estimation based on photoplethysmography (PPG) signal analysis~\cite{ppgPrinciple} with motion compensation, i.e., chest motion analysis, to extract vital signs reliably under real-world conditions.
Experimental results from prototype tablet and vehicle-based systems~\cite{islamPaper1,islamPaper2} demonstrate improved robustness and accuracy in the heart rate and respiratory rate measurements when depth information from a near-infrared time-of-flight camera is incorporated, with comparable performance observed across different skin tones, genders, and age groups.
Beyond measurement accuracy, this effort emphasizes system-level impact by targeting \textit{reductions in hospital length of stay} and enabling \textit{non-disruptive monitoring without extensive wiring} or specialized infrastructure. 

Beyond imaging-based approaches, a representative line of work discussed in the workshop focused on physiological signal sensing and intelligent neural processing systems-on-chip (SoCs) for wearable and implantable medical devices. This line of work targets applications where continuous bio-signal acquisition and real-time inference must be performed under stringent constraints on power consumption, form factor, and data transmission, with demonstrated operation at sub-milliwatt (\textless 1 mW) power levels, making off-chip processing impractical. 
By tightly integrating mixed-signal front ends, embedded feature extraction, and neural network inference directly on-chip, the system enables low-latency classification of physiological signals such as electromyography (EMG), electroencephalography (EEG), electrocardiography (ECG), and neural recordings while operating within strict energy budgets.
One line of work addresses \textit{biomedical rehabilitation}, with a focus on real-time implantable motion intent AI chips for prosthetics~\cite{guPaper1,guPaper2,guPaper3,guPaper4}. These systems integrate multi-channel analog front ends with on-chip neural inference to decode muscle and neural activity directly at the device, enabling responsive prosthetic control while minimizing wireless communication and power consumption. Techniques such as adaptive sampling, dynamic duty cycling, and lightweight neural models allow the system to balance accuracy and lifetime constraints, supporting long-term use in rehabilitation settings. 
Another focuses on intelligent neural interface chips for human-machine interaction, including \textit{human activity and emotion tracking for VR and AR} systems~\cite{guPaper5,guPaper7}.
In this setting, EEG signals are processed on-chip to infer user intent, mental imagery, and affective states without reliance on cameras or external computing platforms. By transmitting compact inference results rather than raw signals, these systems reduce communication overhead and enable portable, camera-free neural interfaces for assistive technologies and clinical monitoring.
Beyond these deployed demonstrations, the workshop discussion highlighted promising directions enabled by intelligent neural processing platforms, including adaptive and personalized bio-signal processing, multi-node body-area multi-chip collaboration~\cite{guPaper8}, and long-term monitoring across diverse use cases. 


A third representative effort discussed in the workshop focused on \textbf{bidirectional brain-machine interfaces}, highlighting the challenges of embedding algorithms and circuits into implantable neural systems under extreme constraints on power, size, thermal safety, and scalability. 
This work centers on the {monolithic integration of neural stimulation and recording} 
supporting closed-loop therapies for neurological disorders such as epilepsy, paralysis, Parkinson's disease, and depression.
The first component addressed high-resolution optical stimulation interface circuits for optogenetics~\cite{yoonPaper1,yoonPaper2,yoonPaper3,yoonPaper4,yoonPaper5}. 
The presented systems monolithically integrate microscopic light-emitting diodes (µLEDs) with silicon neural probes, achieving stimulation sites with dimensions comparable to individual neuron somata. Representative designs integrate tens to hundreds of µLEDs and recording sites across multi-shank probes, enabling spatially confined optical stimulation with nanowatt- to microwatt-level optical power while maintaining tight thermal control. Demonstrations in freely moving and anesthetized animal models showed precise spatiotemporal control of neural activity, including independent stimulation of neurons separated by tens of micrometers and differential control of somatic and dendritic compartments. Custom LED driver ASICs provide fine-grained, per-channel control of stimulation intensity and timing, supporting fast update rates and high radiant-flux resolution required for closed-loop operation.
The second component focused on scalable neural recording circuits~\cite{yoonPaper6,yoonPaper7,yoonPaper8,yoonPaper9} tightly coupled with optical stimulation.
High-density, low-noise recording interfaces were co-designed with stimulation hardware to enable reliable electrophysiological sensing during active optogenetic modulation. The work addressed stimulation-induced artifacts through circuit-level modeling, layout strategies, and embedded signal processing, enabling simultaneous recording of action potentials and local field potentials with artifact amplitudes constrained to clinically relevant levels. Integrated headstage systems combining optoelectrodes, custom stimulation ASICs, and multi-channel recording front ends demonstrate compact form factors suitable for chronic implantation, while supporting large channel counts and real-time operation. 
The second component focused on scalable neural recording circuits~\cite{yoonPaper6,yoonPaper7,yoonPaper8,yoonPaper9} designed for high-channel-count, closed-loop brain interfaces under strict energy and area constraints. 
The presented work introduced modular analog front-end (AFE) architectures~\cite{yoonPaper6} that exploit the spectral characteristics and spatiotemporal structure of neural signals to improve efficiency at scale.
Techniques such as spectrum equalization and continuous-time $\Delta\Sigma$-based signal acquisition compress neural signal dynamic range at the analog front end, enabling high signal fidelity while minimizing energy and area overhead. 
To further reduce power consumption associated with data transmission, the systems embed lossless, signal-aware compression~\cite{yoonPaper7,yoonPaper8} that leverages spatiotemporal correlation in local field potentials and sparsity in action potentials, achieving multi-fold reductions in data rate and substantially lowering dynamic power dissipation without sacrificing recording performance. 
The work also integrates energy-efficient wireless transceivers for bidirectional communication~\cite{yoonPaper9}, enabling simultaneous neural recording and optical stimulation in closed-loop operation without reliance on power-hungry interference cancellation. 
Collectively, these advances demonstrate how tightly coupled circuit architectures, embedded algorithms, and communication strategies are essential for scaling neural recording systems toward dense, long-term, and clinically relevant bidirectional brain interfaces.

A fourth representative effort
explores the full-stack integration of wearable, implantable, and edge-AI-driven systems to enable continuous longitudinal monitoring. This line of works demonstrates the power of multimodal sensing across a variety of clinical needs, including smartwatch-based stress detection~\cite{fazeli2022passive} and the use of self-supervised learning pipelines to manage the complexity of real-world multimodal data~\cite{fazeli2023self}. Beyond traditional vitals, this research extends to specialized medical procedures, such as nondestructive mechanical probing to improve the accuracy of epidural needle placement~\cite{simpson2022epidural}. System-level mobile health platforms are central to this effort, enabling the fusion of personal and environmental data at scale. Platforms such as Raproto~\cite{hamid2024raproto}, Lumos~\cite{watson2023lumos}, and VitalCore~\cite{choi2021vitalcore} provide the necessary infrastructure for real-time health assessment and large-scale longitudinal studies. These platforms have been successfully deployed for pediatric asthma risk prediction by correlating wearable sensor data with environmental factors~\cite{bui2020biomedical, hao2022daily}. The scope of this work also encompasses digital therapeutics and rehabilitation. By leveraging smart insoles and IMU-based gait analysis systems~\cite{ershadi2021gaitoe, grant2023smart}, researchers are developing closed-loop interventions that provide immediate feedback to patients during recovery. Finally, at the implantable level, this line of research demonstrates the successful integration of on-device intelligence for cardiac monitoring and arrhythmia detection~\cite{jia2021enabling, lu2022icd}, ensuring that life-critical analysis occurs directly on the device with high reliability.


An emerging topic highlighted in the workshop is the development of \textbf{personalized cardiovascular digital twins}~\cite{jafariPaper1}. 
This direction envisions continuously updated virtual representation of an individual patient's heart and circulatory system that integrate multimodal clinical data, medical imaging, and wearable sensor measurements. 
Unlike traditional care that relies on episodic checkups and population-level averages, digital twins aim to support predictive, proactive, and individualized care by estimating current physiological state, forecasting disease progression, and enabling simulation of ``what-if'' treatment scenarios to inform clinical decisions. 
Workshop discussions emphasized that effective cardiovascular digital twins require hierarchical physiological models that combine physics-informed, mechanistic representations spanning cellular- and tissue-level processes through organ- and system-level dynamics with AI and machine learning techniques. This hybrid approach provides interpretability and safety while enabling personalization, scalability, and real-time adaptation.
Participants also identified key challenges that currently limit this vision, including limited personalization, incomplete continuous sensing, high computational cost, and the lack of large-scale clinical validation. Despite these challenges, digital twins were viewed as a unifying framework that could transform cardiovascular care from reactive and intermittent monitoring to continuous, predictive, and precision-guided management, representing an important future direction for wearable and implantable medical systems.

\subsection{Key Challenges and Showstoppers}
Discussions during panels and roundtable discussions identified several key challenges and showstoppers that continue to limit the development, scalability, and long-term deployment of wearable and implantable medical systems, particularly those requiring continuous sensing, embedded intelligence, and closed-loop operation under strict safety and energy constraints.

\begin{itemize}
\item \textbf{Trade-offs between monolithic and hybrid integration in implantable neural systems.}
Workshop discussions highlighted a persistent challenge in choosing between monolithic integration of circuits and probes (e.g., Neuropixels~\cite{steinmetz2021neuropixels2}), which reduces interconnect complexity and improves scalability, and hybrid integration (e.g., Neuralink~\cite{musk2019integrated}), which offers lower manufacturing cost, reusability, and flexibility in probe configuration. 
Monolithic approaches enable high channel counts but suffer from high fabrication cost and limited compatibility with flexible, non-silicon probes, while hybrid approaches introduce high lead counts. 
These competing constraints complicate manufacturability, reliability, and cost-effective scaling, posing practical challenges for large-scale clinical deployment of bidirectional brain interfaces.

\item\textbf{Transmit Data vs. Information; Edge vs. Cloud Computing.}
High-channel-count wearable and implantable systems generate large volumes of physiological data, making raw data transmission a major source of power consumption. Workshop discussions emphasized the need to decide what information should be extracted on-device and what should be transmitted, as well as where computation should occur. 
Edge processing can reduce latency, power, and privacy risks, while cloud computing supports more complex models and easier updates. Choosing the right balance between on-device processing and cloud-based computation remains a central challenge for closed-loop and safety-critical medical systems.

\item \textbf{Algorithm upgradeability, programmability, and personalization over time.}
Workshop discussions highlighted the difficulty of enabling algorithm updates in long-lived wearable and implantable medical systems. 
Supporting on-the-fly upgrades to incorporate the most recent algorithms, programmable customization of treatment, and retraining or tuning models using patient-specific data remains challenging under strict constraints on safety, power, memory, and reliability. 
These limitations hinder personalization and long-term adaptation, which are critical for chronic-use and implantable systems that must remain effective as patient conditions evolve.

\item \textbf{Data availability, privacy, and regulatory constraints for AI-driven medicine.}
Discussions highlighted the tension between the large volumes of data required for modern AI and the strict privacy, security, and regulatory constraints governing medical data. 
Challenges include how patient data should be collected, stored, anonymized, and shared; 
whether and how cloud-based computation, e.g., LLM services, can be safely used; 
and how to leverage public or pre-trained AI models while remaining compliant with the Health Insurance Portability and Accountability Act (HIPAA)~\cite{hippa} and related regulations. 
These constraints complicate data aggregation, model training, and cross-institutional collaboration.

\item \textbf{Sensor proliferation vs. usability and adoption.}
The panel emphasized that although a wide range of contact and contactless sensors is becoming available for measuring heart rate, respiration, blood oxygenation, blood pressure, and other physiological parameters, usability and convenience remain critical barriers. 
Non-intrusive, low-burden sensing is essential for sustained adoption, particularly for chronic monitoring. 

\end{itemize}

\subsection{Cross-Cutting Themes and Research Gaps}

\begin{itemize}
    \item \textbf{Insufficient support for long-term programmability and upgradeability.}
    Existing wearable and implantable platforms lack safe, systematic mechanisms for algorithm upgrades, retraining, and personalization over long-term lifecycles spanning years or even decades, while respecting constraints on power, memory, reliability, and regulatory compliance.

    \item \textbf{Limited methodologies for learning from deployment data under privacy and regulatory constraints.}
    Participants noted that while deployed devices continuously collect data, real-world medical data is highly heterogeneous and often weakly labeled. 
    At the same time, strict privacy and regulatory requirements limit data sharing and centralized aggregation, creating a gap in learning frameworks that can leverage such data safely and compliantly.
    
    \item \textbf{Inadequate testing, debugging, and validation for in-body systems}
    Current testing approaches often rely on simplified electrical models, such as resistors or inductors, that fail to capture realistic in-body conditions. 
    As discussed in the breakout sessions, these abstractions do not reflect tissue motion, physiological variability, or complex device-body interactions, making it difficult to anticipate failures before deployment. 
    For example, testing hardware for implantable or swallowable devices remains challenging because the human body is not stationary, and real-world operating conditions cannot be faithfully reproduced in benchtop simulations. 
    This gap increases risk for long-lived, safety-critical devices, where errors discovered after deployment are costly or infeasible to correct.

    \item \textbf{Lack of design methodologies that incorporate usability and quality-of-service guarantees.}
    Participants emphasized that although many sensing modalities are available, sustained adoption depends on usability and minimal burden to patients. Examples included the need for non-intrusive sensing and the importance of maintaining acceptable performance even as batteries degrade, rather than allowing devices to fail abruptly.

\end{itemize}

\subsection{Future Directions and Recommendations to NSF}

{NSF can shape the future of wearable and implantable computing by prioritizing longevity- and adaptation-first design paradigms that support safe in-situ learning and personalization over multi-year lifecycles. Investments in learning methods robust to limited labels and regulatory constraints, together with rigorous testing, validation, and usability- and reliability-aware design tools such as realistic in-body models and digital twins, are essential to ensure safety and dependability in long-lived deployments. Finally, NSF is uniquely positioned to catalyze foundational research on hierarchical, physics-informed digital twins that integrate mechanistic physiology with AI to enable interpretable, predictive, and personalized medicine.}

\begin{itemize}

    \item \textbf{Promote longevity- and adaptation-first design paradigms.}
    Wearable and implantable devices differ fundamentally from short-lived computing systems, yet current research often assumes static behavior or short deployment horizons.
    NSF could support research on safe mechanisms for algorithm updates, in-situ retraining, and personalization over multi-year lifecycles. This would allow devices to adapt to evolving patient conditions without requiring invasive intervention or repeated regulatory reapproval.
    
    \item \textbf{Promote learning methods robust to limited labels and regulatory constraints.}
    NSF could invest in unsupervised, self-supervised, and reinforcement learning approaches, along with privacy-aware techniques, that operate effectively on heterogeneous, weakly labeled deployment data and support continual adaptation through interaction-based feedback, while respecting medical privacy and regulatory requirements.

    \item \textbf{Invest in testing, validation, and usability- and reliability-aware design tools for in-body systems.}
    NSF could support research on realistic in-body models, digital twins, and pre-deployment evaluation methodologies that better capture tissue motion, physiological variability, and device-body interactions, while prioritizing usability-aware, quality-of-service-aware, and graceful-degradation-aware system design as first-class objectives. 
    Together, these efforts would reduce risk and ensure that wearable and implantable devices remain functional, safe, and dependable over long-lived, safety-critical deployments as components age or energy resources diminish.

    \item \textbf{Catalyze research on hierarchical, physics-informed digital twins for personalized medicine.}
    Workshop discussions highlighted digital twins as a promising and unifying framework for transforming medical care from reactive and episodic to continuous, predictive, and individualized. 
    NSF is uniquely positioned to support foundational research on hierarchical physiological models that integrate physics-informed, mechanistic representations across cellular, tissue, organ, and system levels with AI and machine learning. 
    Investments in this area could advance interpretable, safety-aware, and computationally efficient digital twin frameworks that leverage wearable and implantable sensing while enabling personalization, scalability, and real-time adaptation. Such efforts would complement NIH-supported clinical validation by de-risking core modeling, systems, and computational challenges at early stages.
    
\end{itemize}

\section{Research Theme 3: Home ICU, Hospital Systems, and Elderly Care
}
\label{sec:homeICU}

{\subsection{Overview and Motivation}}

The center of gravity for intensive care and chronic disease management is shifting from the clinical ward to the residential living room. Driven by an aging global population, rising healthcare costs, and the desire for ``aging in place,'' the concepts of the \emph{Home ICU} and \emph{hospital-at-home} have emerged as essential pillars of future healthcare. However, the home is a radically different computing environment than the controlled hospital setting and it is characterized by ``noisy'' data, infrastructure instability, and complex social determinants of health that often outweigh purely physiological factors.

\noindent\underline{From Vital Signs to Behavioral and Mental Context.} As it is demonstrated by the efforts discussed in Section~\ref{sec: effort}, effective home care requires a multi-modal approach that extends beyond traditional vitals. Success in this domain involves inferring the \emph{digital biomarkers of daily life}, ranging from the acoustics of joint health to the social mobility patterns that signal escalating loneliness or cognitive decline. A primary challenge identified in the workshop is the \emph{ground-truth gap} for mental and emotional health; currently, measuring these states relies almost exclusively on subjective questionnaires and self-reports, which lack temporal resolution and objectivity. Consequently, the field is moving toward hybrid sensing ecosystems that combine high-fidelity wearables with unobtrusive, battery-free environmental sensors to transform these subjective experiences into objective, scalable, and continuous data streams.

\noindent\underline{The Challenge of the ``Uncontrolled'' Environment.} The transition to the home introduces unique ``showstoppers'' that laboratory-based research often ignores. Technical performance is frequently undermined by infrastructure limitations, such as unreliable broadband and aging housing, and socio-technical barriers, including varying levels of digital and graph literacy. Furthermore, the workshop emphasized that detection and prediction are no longer enough. The ultimate hurdle is identifying the appropriate interventions, determining exactly what actions should follow a detection to improve long-term outcomes without being intrusive. Unlike in-hospital systems, home-based AI must provide safe and accountable feedback that is supportive rather than intrusive, all while navigating a fragmented ecosystem where home-generated data rarely integrates with the clinician’s Electronic Health Record (EHR).

\noindent\underline{A Strategy for Resilient Aging (NSF recommendations).} To bridge these gaps, the NSF is uniquely positioned to lead research that treats the home as a vital engineering constraint. This requires a strategic pivot toward:
\begin{itemize}
    \item \textbf{Objective Ground-Truth Infrastructure:} Developing digital phenotyping and multi-modal frameworks that transform subjective self-reports into objective, scalable measures of mental and social health.
    \item \textbf{Closed-Loop Accountability:} Advancing integrated systems that connect sensing and inference to validated interventions, ensuring that automated coaching or feedback mechanisms are effective, safe, and transparent.
    \item \textbf{Environment-Robust and Literacy-Aware Design:} Creating ``born-to-fail'' systems that maintain utility despite connectivity loss and designing interfaces that match the cognitive and literacy levels of older adults and informal caregivers.
\end{itemize}

By integrating social, behavioral, and clinical context into a unified cyber-physical framework, the next generation of home-based systems will transform from passive monitors into active, trustworthy interventions that preserve independence and improve outcomes for the most vulnerable populations.

\subsection{Representative Existing Efforts and Promising Directions}\label{sec: effort}

This section synthesizes selected examples discussed in the workshop that reflect state-of-the-art strategies and system-level progress in home-centered intensive care, hospital platforms, and technology-enabled elderly care.




One representative effort discussed in the workshop focused on the use of artificial intelligence and integrated technologies to support aging populations, with particular emphasis on home-based care, social context, and human-centered system design. 
This presentation highlights that effective AI systems for older adults must extend beyond physiological sensing to address social isolation, functional decline, and daily living challenges that strongly influence health outcomes. 
The presentation emphasized the growing prevalence of loneliness and social isolation among older adults and their links to increased mortality, cardiovascular risk, and cognitive decline, motivating the need for continuous, in-home monitoring and intervention rather than episodic clinical encounters. 
The work demonstrated community-deployed sensing platforms, iCareLoop~\cite{DemirisPaper1,DemirisPaper2,DemirisPaper3,DemirisPaper4,cho2024perceptions}, that combine passive in-home sensors, questionnaires, and user interfaces with centralized analytics to detect changes in behavior, mobility, sleep, and social engagement, enabling early identification of escalating risk and timely intervention.
The scope of this research further extends to high-stakes pediatric and perinatal care, including advanced neonatal oxygen monitoring~\cite{demauro2025oxygen} and alarm reduction strategies for infants with bronchopulmonary dysplasia~\cite{herrick2022alarm} to mitigate alert fatigue in intensive care, as well as clinical decision-support systems for predicting extubation failure risk and AI-driven analysis of communication patterns in autistic children~\cite{yang2025understanding}. 
Concretely, the system employs a heterogeneous Internet of Things (IoT) sensing suite that includes motion and contact sensors for room-level activity and daily living patterns, bluetooth low energy (BLE) tag-based proximity sensing to infer time spent indoors versus outdoors and social mobility, temperature and humidity sensors to capture environmental context and hygiene-related events (e.g., showering), smart plugs to monitor appliance usage, sleep-mattress sensors that extract multi-metric sleep quality features, and optional wearable activity trackers for step counts and physical activity. 
These devices stream data to an in-home gateway that aggregates and securely uploads data to a cloud-backed management platform, where time-series features are computed and analyzed using anomaly detection and machine learning models. The resulting risk indicators are surfaced through dashboards and clinician-in-the-loop decision support, forming a closed-loop pipeline that links continuous sensing, behavioral inference, and personalized interventions.

In addition, the talk discussed data-driven approaches for assessing fall risk in aging populations, including FRED: Fall Risk Evaluation Database~\cite{DemirisPaper5}, which is an open-source database constructed from large-scale electronic health record (EHR) data (derived from MIMIC-III~\cite{johnson2016mimic}) that organizes demographic information, longitudinal Morse Fall Scale assessments, time-stamped clinical notes, and administered medication records into labeled risk intervals, enabling data-driven modeling and prediction of changes in fall risk over time. 
The presented work illustrates how intelligent home and hospital systems can extend clinical visibility into daily living environments and support aging in place through continuous monitoring and data-informed care delivery.


Another representative effort discussed in the workshop focused on wearable acoustic and vibration sensing combined with machine learning to enable continuous, non-invasive monitoring of physiological function in home and clinical settings. 
This work leverages the fact that the human body continuously generates mechanical signals, including cardiac vibrations, respiratory sounds, joint acoustics, and peripheral pulse waveforms. 
By capturing these signals using compact, body-worn sensors, the system enables frequent, low-burden assessment of cardiovascular and musculoskeletal health outside of traditional clinical environments. One example is a \textbf{multimodal wearable sensing} chest patch ~\cite{inanPaper1,inanPaper2,inanPaper3} that integrates ECG, seismocardiography (SCG), PPG, and inertial sensing to estimate hemodynamic parameters such as cardiac output, filling pressures, and congestion status in patients with heart failure, supported by over a decade of clinical validation studies across diverse patient populations.
Beyond \textit{cardiovascular monitoring}, this work also demonstrated sensor-embedded wearables for \textit{joint health and rehabilitation}, such as multimodal sensing knee braces~\cite{inanPaper4, inanPaper5,inanPaper6,inanPaper7,inanPaper8} that combine acoustic emissions, bioimpedance, and motion sensing to assess tissue damage, inflammation, and injury risk during daily activity. 
Machine learning pipelines transform raw vibration and acoustic signals into clinically meaningful metrics, enabling objective, longitudinal tracking of disease progression and recovery. 
These systems illustrate how tightly integrated sensing, signal processing, and learning can support early detection, remote monitoring, and proactive intervention, which are central goals of Home ICU and elderly care models that aim to reduce hospitalizations, enable aging in place, and support continuous care beyond episodic clinic visits.

A third representative effort highlighted in the workshop focuses on the development of adaptive, learning-enabled deep brain stimulation systems~\cite{PajicPaper1,PajicPaper2,PajicPaper3,PajicPaper4,PajicPaper5,PajicPaper6} that transform traditional continuous stimulation into closed-loop, patient-specific neuromodulation. 
Existing commercial DBS devices typically deliver fixed-frequency periodic pulses that, while effective at alleviating motor symptoms, are energy-inefficient, shorten battery lifetime to only a few years, and may induce side effects such as speech impairment. 
These limitations motivate moving beyond fixed, one-size-fits-all stimulation toward adaptive, personalized controllers that can balance therapeutic efficacy with efficiency and safety.
This line of work demonstrates clinically deployed 
adaptive controllers that leverage neural biomarkers to modulate stimulation in real time, reducing power consumption while maintaining therapeutic efficacy in both clinic and home environments~\cite{PajicPaper1}. 
Building on this foundation, the authors cast DBS control as a sequential decision-making problem and introduce learning-based approaches ranging from 
deep reinforcement learning for synthesizing personalized stimulation patterns~\cite{PajicPaper3} to more sample-efficient contextual bandit methods~\cite{PajicPaper2}, 
that enable real-time execution on resource-constrained embedded devices. 
Recognizing the safety and regulatory barriers to online experimentation with patients, they further develop offline reinforcement learning frameworks that train controllers using historical clinical data~\cite{PajicPaper5,PajicPaper4}, together with model-based offline policy evaluation techniques that estimate controller performance without direct \emph{in vivo} testing and extensions that account for sparse human feedback signals when assessing patient-centered outcomes~\cite{PajicPaper6}. 
These contributions establish an end-to-end cyber-physical pipeline spanning physiological modeling, learning-based control, offline training, and safety-aware evaluation, advancing adaptive neuromodulation toward clinically practical and trustworthy closed-loop therapies.

A promising direction identified in the workshop is wireless at-home medical sensing that leverages signals of opportunity and \textbf{battery-free} technologies to enable scalable, low-cost health monitoring. 
By repurposing everyday radio frequency, acoustic, and near field signals as opportunistic sensing modalities, these systems support continuous, passive health tracking without bulky wearables, frequent charging, or user burden, making long term deployment in home environments both practical and sustainable.
Rather than relying on dedicated, battery-powered devices, this work demonstrates battery-free and minimally obtrusive sensing modalities built on passive RFID and NFC technologies, enabling scalable deployment in home environments. 
Examples include flexible on-skin RFID ``tattoos"~\cite{KumarPaper1} that detect subtle muscle activity and skin stretch through resonant frequency shifts of the antenna (e.g., millimeter-scale deformation producing measurable MHz-level changes), supporting applications such as muscle atrophy monitoring and speech recognition for individuals with voice disabilities. 
The approach further extends to ingestible RFID sensor capsules for esophageal monitoring~\cite{KumarPaper2} and acoustic sensing through everyday devices such as sonic toothbrushes~\cite{KumarPaper3} to detect dental health issues, illustrating how commodity wireless and acoustic infrastructures can double as medical sensing platforms. 
These systems highlight a deployment-friendly paradigm that emphasizes battery-free operation, low cost, and seamless integration into daily life, enabling continuous home health monitoring without imposing additional burden on patients or caregivers.

Another promising direction is contactless health sensing that couples wireless sensing with machine learning to extract digital biomarkers without requiring cameras or body-worn devices. 
By leveraging \textbf{ambient RF reflections}~\cite{ZhaoPaper1}, these systems infer human pose, activity, and motion, even through walls, while preserving privacy and reducing instrumentation requirements, making them well-suited for residential and hospital-at-home settings. 
Beyond activity recognition, wireless signals can capture fine-grained physiological measurements such as respiration, heartbeats, and sleep stages~\cite{ZhaoPaper2}, with performance approaching clinical sleep-lab standards, enabling continuous overnight monitoring without electrodes or wearables. 
The framework further supports AI-driven assessment of medication adherence, daily living patterns, and cognitive or mental states through smart-glasses-based~\cite{ZhaoPaper3} and environmental sensing, broadening the scope from vital signs to functional and behavioral health. 
Together, these advances suggest a future in which unobtrusive, infrastructure-based sensing provides longitudinal, camera-free monitoring that seamlessly integrates into everyday environments, enabling reliable and privacy-aware care outside traditional clinical settings.

\subsection{Key Challenges and Showstoppers}

The transition of intensive and elderly care from clinical environments to the home represents a paradigm shift in healthcare delivery, powered by ubiquitous sensing and AI. However, the realization of this vision is currently obstructed by ``showstoppers'' such as infrastructure instability, fragmented clinical workflows, and a critical lack of objective ground truth for mental and emotional health monitoring. Addressing these hurdles requires a holistic approach that moves beyond mere detection toward safe, interoperable interventions that are synchronized with the social and operational realities of domestic life.

\begin{itemize}
    \item \textbf{Lack of ground truth for mental, emotional, and social health sensing.}
    A fundamental challenge in monitoring mental, emotional, and social well-being is the lack of reliable ground truth. Current systems often rely on subjective self-reports or surveys to label mental states, which limits objectivity, temporal resolution, and scalability. While sensors can capture behaviors related to loneliness and daily activities, such as routines or hygiene-related actions, mapping these observations to validated mental or emotional states remains difficult without robust reference signals. 

    \item \textbf{Lack of effective and safe interventions beyond detection and prediction.}
    Beyond sensing and detection, designing effective and acceptable interventions remains an unresolved challenge. 
    This includes determining what actions should follow detection, how to provide feedback that is supportive rather than intrusive, and how to ensure safety and accountability, particularly when using automated or AI-driven coaching systems. 
    Additional challenges include evaluating effectiveness in social cyber-physical systems where feedback is less controlled, ensuring interventions are acceptable to older adults, and moving beyond detection and prediction toward interventions that demonstrably improve long-term outcomes.

    \item \textbf{Infrastructure and connectivity limitations in home environments.}
    Workshop discussions emphasized that many home and elderly care settings lack the reliable broadband, low-latency connectivity, and power stability assumed by modern AI and sensing systems. 
    Intermittent Internet access, wireless interference, and aging housing infrastructure directly degrade data completeness, real-time inference, and remote monitoring capabilities. 
    Without dependable connectivity and basic infrastructure, even well-designed sensing and AI solutions cannot operate reliably, creating a fundamental barrier to deployment at scale.

    \item \textbf{Socio-technical barriers to adoption in home and elderly care settings.}
    Workshop discussions emphasized that the effectiveness of AI-enabled home ICU and elderly care systems is fundamentally constrained by socio-technical factors beyond algorithmic performance. 
    Key challenges include limited availability and reliability of broadband Internet, gaps in digital and graph literacy, varying levels of familiarity and comfort with technology, and wide variability in housing conditions and living environments. 
    These factors directly affect system usability, trust, data quality, and sustained adoption. 
    Participants noted that developing AI innovations that are responsive to social needs in the broadest possible manner is not only a social and ethical imperative, but also an economic one~\cite{DemirisMentionWork1}, as these barriers strongly influence scalability, sustainability, and real-world impact.

    \item \textbf{Fragmented care ecosystems and lack of interoperability with clinical workflows.}
    Home sensing platforms often operate independently from electronic health records, hospital IT systems, and established clinical workflows. 
    This fragmentation creates data silos, duplicate documentation burdens, and limited clinician visibility into home-generated data, reducing the clinical utility of otherwise rich sensing streams. 
    Without standards-based integration, interoperability, and workflow alignment, home monitoring systems remain peripheral tools rather than actionable components of care delivery.

    \item \textbf{Operational and maintenance burden at scale.}
    Long-term home deployments introduce significant logistical challenges, including device installation, battery replacement, troubleshooting, software updates, and user support. 
    These operational costs and maintenance demands are often underestimated and become prohibitive when scaled across large populations. Systems that require frequent intervention or technical support are unlikely to be sustainable for caregivers, healthcare providers, or community organizations.

\end{itemize}

\subsection{Cross-Cutting Themes and Research Gaps}

Effective home-based care requires a transition from idealized laboratory models to systems that are resilient to the inherent unpredictability of domestic life and the complexities of human emotion. Current research gaps highlight a critical need for holistic frameworks that move beyond simple physiological detection toward objective mental health ground truths and validated, non-intrusive interventions. Ultimately, these systems must unify environmental robustness and user-centric design with a deep integration of social determinants to support sustainable aging-in-place.

\begin{itemize}
\item \textbf{From subjective self-reports to objective, scalable ground truth for mental and social health.} This challenge raises open questions about what sensing modalities are appropriate for studying mental health, how emotional states can be inferred from behavioral or physiological data, how contextual factors that individuals may not explicitly report or even recognize can be accounted for, 
    and whether more objective, continuous measures can be developed to complement or replace questionnaires. 
    Addressing these gaps is critical for building trustworthy models that generalize across individuals and contexts.
    \item \textbf{From detection and prediction to validated and effective interventions.}
    Even when mental or emotional states can be detected or predicted, there is a significant research gap in defining and validating what actions should follow. 
    Current efforts exploring intervention mechanisms, such as technology-mediated coaching agents, remain limited in evidence regarding appropriateness, acceptability, and effectiveness, particularly for older adults. 
    Open questions persist around how to design interventions that are timely and supportive without being intrusive, how to incorporate safeguards against undesirable outcomes, and how to evaluate effectiveness in real-world settings. 
    More broadly, there is a lack of rigorous evaluation frameworks and outcome metrics beyond detection accuracy, hindering progress toward interventions that reliably improve long-term mental, emotional, and social health outcomes.

    \item \textbf{Lack of environment-robust models for home and elderly care environments.}
    Existing AI and sensing systems are often designed and evaluated under idealized assumptions about connectivity, housing conditions, and user behavior. There is a lack of models, benchmarks, and datasets that capture real-world variability in home environments, including intermittent connectivity, heterogeneous housing layouts, and non-uniform sensor placement. This gap limits the ability to predict system performance and reliability outside controlled study settings.

    \item \textbf{Insufficient integration of usability, literacy, and trust into system design.}
    Current research largely treats digital literacy, graph literacy, and user familiarity as external adoption factors rather than first-class design constraints. Few systems explicitly model how older adults, caregivers, and clinicians interpret system outputs, respond to alerts, or build trust over time. As a result, technically capable systems often fail to translate into sustained use.

    \item \textbf{Fragmented approaches to integrating social and clinical context.}
    Many existing efforts focus on physiological sensing or clinical risk prediction in isolation, with limited incorporation of social factors such as isolation, caregiving availability, and daily living patterns. This separation constrains the development of holistic systems that can support aging in place and hospital-at-home care in a realistic manner.

\end{itemize}

\subsection{Future Directions and Recommendations to NSF}

To bridge the gap between clinical potential and domestic reality, the following recommendations outline a strategic roadmap for federal research centered on resilience and accountability. These priorities emphasize a shift toward objective ground-truth mechanisms for mental health and the development of safe, closed-loop systems that move beyond mere detection to validated interventions. By aligning technological innovation with real-world infrastructure, user literacy, and integrated care workflows, these initiatives aim to catalyze the next generation of sustainable, sociotechnical aging-in-place solutions.

\begin{itemize}

    \item \textbf{Invest in objective, ground truth, and validation infrastructure.}
    NSF should support research that develops objective, continuous, and scalable ground-truth mechanisms for mental, emotional, and social health, moving beyond reliance on subjective self-reports.
    This includes investment in digital phenotyping approaches that leverage passively sensed smartphone data, such as device usage logs and proxy measures of socialization, as well as multimodal physiological and behavioral signals derived from audio, video, and other sensing modalities. 

    \item \textbf{Support safe, accountable systems integrating sensing, inference, and intervention.}
    NSF should support research that moves beyond isolated advances in sensing or prediction toward integrated, closed-loop systems that explicitly connect detection with intervention. 
    This includes encouraging projects that co-design sensing, modeling, and intervention components and evaluate their combined impact on real-world outcomes rather than detection accuracy alone. 
    Priority should be given to research on designing, deploying, and validating interventions that are effective, acceptable, and safe, particularly for older adults and other vulnerable populations, with explicit mechanisms for safeguards, accountability, and appropriate human oversight. 
    Interdisciplinary collaborations spanning engineering, social sciences, and health domains will be essential to ensure that such systems deliver timely, supportive feedback and interventions without being intrusive.

    \item \textbf{Invest in environment-robust models, benchmarks, and real-world testbeds.}
    NSF should prioritize research that evaluates AI and sensing systems under realistic home operating conditions rather than controlled laboratory settings. 
    Programs should fund multi-site living-lab deployments across diverse housing types and socioeconomic contexts, and support open datasets and benchmarks that explicitly capture intermittent connectivity, heterogeneous layouts, sensor placement variability, and long-term maintenance challenges. 
    In addition to accuracy, proposals should report robustness, reliability, uptime, and fault tolerance metrics to ensure systems generalize beyond idealized environments and remain dependable in everyday home care settings.

    \item 
    {\textbf{Advance literacy-aware AI system design.}}
    NSF should treat human factors, including digital and graph literacy, as core technical constraints rather than downstream adoption concerns by encouraging tightly integrated collaborations among HCI researchers, clinicians, gerontologists, and systems engineers. 
    Projects should incorporate co-design with older adults and caregivers, develop interpretable and explainable interfaces that match users' cognitive and literacy levels, and evaluate long-term engagement, trust formation, and alert fatigue through longitudinal studies. 
    Funding mechanisms should incentivize adaptive interfaces and standardized usability and trust metrics so that systems are not only technically capable but also usable, understandable, and sustainable in practice.

   { \item \textbf{Enable environment-embedded, battery-free, and contactless sensing to complement wearable technologies for home health monitoring.} NSF should prioritize research into sensing paradigms that embed intelligence directly into the domestic environment via ambient RF, acoustic, and near-field signals. These contactless modalities are designed to complement, rather than replace, wearable technologies by providing continuous, ``always-on'' background coverage without requiring active user compliance or frequent battery maintenance. While wearables remain the gold standard for high-fidelity, personalized physiological tracking, environment-embedded systems, such as passive backscatter tags and device-free radios, fill critical gaps by capturing motion, sleep, and vital patterns when on-body sensors are being charged or are intentionally removed. By advancing research in passive hardware and privacy-preserving inference, NSF can enable a dual-layered sensing architecture where environmental and wearable data are fused to improve overall system robustness and reduce the logistical burden on elderly users. This integrated approach ensures that aging-in-place and hospital-at-home care models remain dependable through a combination of high-fidelity wearable measurements and resilient, infrastructure-based monitoring.}
    
    \item \textbf{Promote socio-technical integration of social, behavioral, and clinical context.}
    NSF should encourage research that moves beyond siloed physiological monitoring toward multimodal frameworks that jointly model environmental, behavioral, social, and clinical signals. 
    Programs should support data fusion methods, shared datasets linking home sensing with EHR and care outcomes, and decision-support tools that incorporate social determinants such as isolation, caregiving availability, and daily living patterns. 
    Emphasis should be placed on closed-loop systems that connect sensing, inference, and personalized interventions, enabling proactive and comprehensive support for aging in place and hospital-at-home care.

\end{itemize}

\section{Research Theme 4: Medical Sensing, Imaging, and Reconstruction
}
\label{sec:medicalSensing}

\subsection{Overview and Motivation}

Medical imaging and sensing are the ``eyes'' of modern medicine, providing the fundamental data required for diagnosis, intervention, and longitudinal care. However, the field is currently caught between the promise of high-resolution, AI-driven insights and the harsh reality of computational and data bottlenecks. As imaging modalities move toward micrometer-scale 3D and 4D datasets, the traditional ``linear'' pipeline, where hardware captures data and software later reconstructs it, is becoming insufficient. We are entering an era that requires a fundamental reimagining of the imaging architecture, shifting from rigid, fixed-function scanners to modular, software-defined systems that are deeply integrated with AI.

\noindent\underline{The Efficiency and Trust Paradox.} As discussed in the workshop, the current state of the art is defined by two primary tensions: annotation scarcity and clinical reliability. While deep learning has revolutionized image reconstruction, the ``data hunger'' of these models is a major showstopper; expert clinicians cannot spend hundreds of hours annotating a single micro-CT volume. This has motivated representative efforts in annotation-efficient learning and the use of foundation models such as SAM to ``bootstrap'' medical intelligence. Simultaneously, to move beyond laboratory success, systems must adopt expert-in-the-loop paradigms, such as \emph{Just-Enough Interaction}, ensuring that AI serves as a tool for clinical quantitative analysis rather than a ``black box'' that clinicians cannot correct or trust.

\noindent\underline{Hardware as a Clinical Bottleneck.} The workshop highlighted a critical \emph{memory wall}: modern medical DNNs are outgrowing the 16-24~GB capacities of standard GPUs, leading to significant performance penalties during the training of large foundation models. This mismatch underscores the need for hardware-aware model design and the exploration of non-classical paradigms, such as quantum computing, to handle the exponential complexity of 4D reconstruction and optimization. Furthermore, the move toward synthetic data generation and generative anonymization offers a promising pathway to bypass privacy constraints, allowing for the sharing of clinically rich motion data (e.g., stroke screening) without compromising patient identity.

\noindent\uline{A Strategy for Practical Virtual-Physical Synergy in Medical Imaging (NSF Recommendations).} To accelerate the translation of these technologies, the NSF is positioned to foster a new ecosystem of \emph{Hybrid Virtual-Physical Imaging}. This strategy involves:
\begin{itemize}
    \item \textbf{Digital Twins and Healthcare Avatars:} Creating open-source, interoperable ``scanner-patient'' twins to simulate and validate imaging protocols before they ever touch a physical patient.
    \item \textbf{Modular, Scalable Architectures:} Moving toward ``software-defined'' scanners that can reallocate compute resources (e.g., GPU, FPGA, TPU) dynamically based on the clinical task and product tier.
    \item \textbf{Multi-Task Collaborative Learning:} Supporting AI frameworks that do not merely segment an image in isolation but integrate detection, diagnosis, and treatment planning into a unified, clinically-aware model.
\end{itemize}

By treating the imaging pipeline as a unified cyber-physical system, where the sensor, the reconstruction algorithm, and the clinical workflow are co-designed, we can lower the cost of innovation and ensure that high-fidelity medical imaging is accessible across all care settings, from high-end radiology suites to smartphone-based screenings in resource-constrained environments.

\subsection{Representative Existing Efforts and Promising Directions}

This section summarizes representative efforts discussed in the workshop that demonstrate how AI-based medical sensing, imaging, and reconstruction systems are beginning to address fundamental challenges across the entire pipeline. In particular, current research prioritizes annotation-efficient learning strategies and the use of foundation models to achieve high-performance segmentation and classification with substantially reduced expert labeling requirements. To ensure clinical reliability and facilitate real-world translation, these approaches are increasingly coupled with expert-in-the-loop quantitative analysis frameworks and generative anonymization techniques, enabling rapid human correction of automated outputs while supporting privacy-aware data augmentation and sharing in high-stakes clinical settings. Looking forward, the workshop highlighted a broader architectural shift toward modular, software-defined imaging systems and emerging ``medical metaverses,'' in which digital twins are used to simulate, optimize, and validate imaging protocols and system resilience prior to physical deployment.

One representative effort discussed in the workshop addressed the annotation challenge in AI-based medical image analysis, a fundamental bottleneck for scalable medical sensing, imaging, and reconstruction. 
This line of work highlights that medical image annotation is expensive, labor-intensive, inconsistent, and often infeasible at scale, particularly for high-dimensional 3D and 4D imaging data with complex anatomical structures. 
For example, micro-CT imaging operates at micrometer-scale resolution, producing extremely large volumetric datasets (e.g., on the order of thousands of voxels per dimension), where annotating a single 3D image can require over a hundred hours of expert effort.
To reduce reliance on exhaustive expert labeling, the presented approaches explore sparse, selective, and representative annotation strategies~\cite{DannyWork1,DannyWork2,DannyWork4} that identify the most informative samples for expert annotation in a one-shot or minimally iterative manner. 
Specifically, these strategies extract unsupervised feature embeddings from 3D image slices, select a small subset of highly representative and diverse slices that minimize redundancy through coverage-score-based optimization, and then apply self-training with pseudo-labels generated by complementary 2D and 3D models to propagate supervision to unlabeled regions.
These methods have been demonstrated in 
applications including 3D segmentation of knee cartilages, bones, and leg muscles~\cite{DannyWork1} in MRI images using sparse annotation, as well as segmentation of cartilage and bone structures in embryonic chondrocranium and dermatocranium~\cite{DannyWork2} from high-resolution 3D micro-CT images,
achieving competitive segmentation performance compared with state-of-the-art deep learning methods while using less than approximately 20\% of annotated data.

Another representative effort discussed in the workshop explores how foundation models can be leveraged to enhance medical image classification~\cite{DannyWork3}.
This work builds on the Segment Anything Model (SAM)~\cite{Kirillov_2023_ICCV}, a segmentation foundation model developed by Meta AI and trained on large-scale image and mask datasets, including medical images. 
The approach uses SAM to generate segmentation and boundary prior maps that emphasize relevant object regions in input images and suppress background content. 
These priors are combined with raw images 
through a model that processes raw and SAM-augmented inputs in parallel and merges their predictions. 
Experiments show that this SAM-based augmentation~\cite{DannyWork3} improves classification accuracy and reliability compared with models trained on raw images alone, achieving state-of-the-art performance on benchmark datasets.

A third representative effort discussed in the workshop focused on quantitative medical image analysis for critical clinical tasks, emphasizing a structured, quality-aware pipeline that combines automation with expert-in-the-loop~\cite{sonkaPaper1} framework. 
Fully automatic medical image segmentation can fail locally in real clinical data due to factors such as pathological changes in appearance, imaging artifacts, low signal-to-noise ratio, unusual anatomy, and scanner variability.
When segmentation fails, clinicians are often forced to manually redraw contours slice by slice, a process that is extremely slow, tedious, and error-prone, and which represents a major barrier to clinical adoption.
The proposed approach decomposes analysis into multiple stages, beginning with object detection as a pre-segmentation step, followed by accurate object segmentation. 
To ensure clinical reliability, automated segmentation quality assessment (SQA) is then applied to identify guaranteed-correct segmentations and to flag potential imperfections that require further attention. 
A key contribution highlighted is the integration of correction mechanisms that preserve global automation while enabling fast, minimal human intervention. 
Specifically, globally optimal graph-based segmentation is combined with a ``just-enough interaction'' (JEI) mechanism that allows experts to correct local errors in seconds rather than redrawing entire structures.
Depending on task criticality and confidence levels, the pipeline supports multiple modes, including expert-guided JEI, SQA-guided JEI, or fully automated correction driven by SQA, with final human approval retained for accountability. 
This expert-in-the-loop paradigm enables accurate, timely, and reproducible quantitative analysis in high-stakes clinical settings, reducing manual effort while preserving safety, trust, and clinical relevance.
Recent extensions further incorporate latent diffusion-based segmentation models~\cite{sonkaPaper2} into the SQA framework. These models operate in low-dimensional latent spaces, significantly reducing memory and computation costs while improving robustness to noise and uncertainty. 
Across multiple medical imaging modalities and datasets, these approaches demonstrate state-of-the-art segmentation accuracy, strong noise resilience, orders-of-magnitude faster inference enabled by requiring only a handful of sampling steps rather than hundreds or thousands, and built-in uncertainty estimation, while keeping clinicians in control through lightweight verification and correction steps.

A fourth representative effort focused on advancing trustworthy radiology systems that can be safely and effectively translated into real-world clinical practice. 
This work addresses the gap between research and deployable clinical systems by emphasizing trustworthiness across the full AI lifecycle, including data quality, model generalizability, interpretability, safety, and fairness. 
The presented efforts span a range of radiology applications, such as risk assessment~\cite{dadsetan2022deep}, triage, diagnosis~\cite{aboutalib2018deep}, prognosis~~\cite{hao2022survivalcnn}, and treatment response prediction, and explicitly account for challenges such as limited and imbalanced data, multi-modality inputs, long-tailed distributions, and cross-site variability. 
Core approaches include learning with imbalanced~\cite{gao2020handling} or imperfect annotations~\cite{hao2020inaccurate}, handling out-of-distribution cases, integrating clinical and biological knowledge into model design, and enabling human-AI collaboration through explainable and clinician-guided workflows. 

A fifth representative effort discussed in the workshop focused on synthetic image and video generation as a means to address data scarcity, privacy constraints, and limited data sharing in medical applications, with a particular emphasis on facial video analysis~\cite{10.1007/978-3-031-96625-5_26,cai2022deepstroke}. 
The work targets scenarios such as emergency stroke screening, where facial motion and expression contain clinically relevant information, but raw patient videos cannot be freely shared. 
To address this, the presented approach uses generative anonymization through cross-identity video motion retargeting, transferring patient motion to synthetic faces to preserve clinically meaningful dynamics while removing identifiable appearance.
Building on this foundation, the work advances toward facial video foundation models using techniques such as joint transformation and synthesis, 3D-aware talking-head generation, and latent flow diffusion models for image-to-video generation, enabling temporally coherent and controllable video synthesis. 
Results demonstrated across multiple datasets show realistic video generation for unseen subjects, improved temporal consistency, and strong performance compared to prior methods, while enabling evaluation along realism, utility, privacy, and diversity dimensions. 

A sixth representative effort discussed in the workshop focused on improving the use of routinely acquired imaging data for adaptive radiotherapy, where image quality and consistency are critical for treatment planning and adjustment. 
This work explores patient-specific learning~\cite{peng2025unsupervised,pan2025cycle} approaches that use limited prior information from each patient to enhance or reconstruct medical images without requiring large paired training datasets. 
The results demonstrate that patient-specific modeling can enable more reliable image-based analysis and decision-making in adaptive treatment workflows, highlighting a promising direction for personalized and data-efficient medical imaging systems.

Complementing these advances in high-end imaging systems and analysis, another representative effort explored how clinically meaningful medical sensing can be achieved using low-cost, widely available consumer hardware, significantly broadening accessibility and scalability.
This work shows that smartphones and inexpensive audio peripherals can be used to conduct diagnostic tests~\cite{chan2022performing,chan2022off}, such as hearing screening, with performance comparable to specialized clinical equipment. By shifting complexity from hardware to software through signal processing and calibration, these systems dramatically reduce cost while maintaining clinical reliability, enabling scalable deployment in resource-constrained and global health settings.

A seventh representative effort
focuses on transforming raw biomedical signals and medical images into accurate, quantitative biomarkers that directly support diagnosis, prognosis, and image-guided therapy. This work bridges the gap between raw data acquisition and clinical action by emphasizing the reliability of the sensing and reconstruction pipeline, enabling automated AI analysis in high-stakes clinical settings through advanced methods for image segmentation quality assessment~\cite{zaman2023patch} and the detection of erroneous regions in automated outputs~\cite{zaman2023segmentation}. The research further advances the efficiency of imaging pipelines via diffusion-based segmentation models~\cite{zaman2024surf, zaman2025latentdiffusionmedicalimage} and fully automated segmentation frameworks~\cite{zhang2025fully}, while also extending into the nanoscale through AI-enhanced nuclear magnetic resonance spectroscopy to improve data quality and acquisition speed~\cite{kong2020artificial}. The high-fidelity latent representations produced during segmentation enable more accurate disease detection and outcome prediction for conditions such as Takotsubo Syndrome (TTS) and non-small cell lung cancer~\cite{roy2024samlfdiag, zaman2024diagnosis, gainey2023predictive}. By emphasizing structured assessment of AI-generated content and the extraction of clinically meaningful features, this work ensures that next-generation imaging systems deliver the precision and reliability required for real-world clinical workflows.

Promising directions discussed in the workshop highlighted the concept of a medical technology and AI metaverse~\cite{JonPaper1} as an enabling framework for next-generation imaging systems. 
This vision centers on the use of digital twins of scanners, patients, and clinical environments to support virtual comparative scanning, raw data sharing, augmented regulatory science, and simulation-driven evaluation of imaging technologies. 
By enabling imaging protocols, reconstruction methods, and analysis pipelines to be tested and optimized in virtual environments before physical deployment, such an approach can reduce development cost, accelerate innovation, and improve robustness and generalizability. 
The discussion emphasized that these \textit{virtual and hybrid environments} can also facilitate large-scale validation, safer regulatory evaluation, and more flexible data sharing, offering a practical pathway to bridge research, clinical translation, and real-world deployment at scale.

Another promising direction highlighted a system-level perspective grounded in practical deployment experience, as practitioners from industry emphasized the importance of designing next-generation medical imaging systems to improve commonality across product models, increase resilience to hardware obsolescence and supply chain, and reduce R\&D and overall system costs. 
The discussion underscored architectural opportunities to reduce data bottlenecks and better support real-time sensor and image data interpretation. 
A key focus was on enabling computational workloads to be scaled and reallocated based on system tier and available computing resources, recognizing that hardware configurations are expected to differ across product tiers. 
Software-defined and modular system designs were highlighted as a means to preserve flexibility, upgradability, and performance across diverse deployment contexts.







\subsection{Key Challenges and Showstoppers}

Training modern medical DNNs is increasingly limited by GPU memory and storage constraints, as model and data requirements often exceed the 16-24 GB capacity of current devices. While offloading techniques can reduce memory pressure, they introduce significant data-movement overheads that degrade performance and restrict model scale and resolution. This growing gap between algorithmic demands and hardware efficiency remains a critical bottleneck for scalable and reproducible medical imaging and reconstruction.

\begin{itemize}
    \item \textbf{Memory, storage, and hardware efficiency as bottlenecks for training large medical DNNs.}
    A key challenge in medical sensing, imaging, and reconstruction is the growing mismatch between the memory and storage requirements of modern deep neural networks and the capabilities of existing GPU-based systems. 
    State-of-the-art medical imaging models, 
    often require tens of gigabytes of memory during training, while even high-end GPUs typically provide only 16-24 GB of on-device memory. 
    In addition, large intermediate activations, model parameters, and training data impose substantial pressure on GPU memory hierarchies and storage subsystems, further limiting scalability. 
    While existing approaches, such as offloading tensors to CPU memory or SSD-based storage~\cite{yang2025agile}, can partially alleviate GPU memory pressure, these techniques often incur significant performance penalties due to increased data movement and reduced training throughput. 
    As a result, training speed and efficiency are frequently sacrificed to fit models within available memory budgets, constraining model complexity, batch size, and resolution.
    At the same time, although specialized hardware accelerators such as GPUs, FPGAs, and TPUs offer substantial computational capability, current medical AI workflows often fail to fully exploit these platforms due to insufficient hardware-aware model design and software support. 
    This gap between algorithmic and hardware efficiency represents a critical showstopper for scalable, efficient, and reproducible training of next-generation medical imaging and reconstruction models, particularly for large foundation models and real-time clinical applications.

\end{itemize}

\subsection{Cross-Cutting Themes and Research Gaps}

This section highlights key research gaps that limit the scalability, robustness, and clinical impact of current AI-based medical sensing, imaging, and reconstruction systems. In particular, existing approaches are constrained by fragmented, task-specific learning paradigms that fail to capture the integrated nature of clinical decision-making, as well as by an overreliance on classical computing architectures that struggle with growing data and computational complexity. Addressing these gaps requires new collaborative learning frameworks and the systematic exploration of emerging computing paradigms that can jointly advance clinical insight integration and computational efficiency.

\begin{itemize}
    \item \textbf{Fragmented task-specific learning limits clinical insight integration.}
    The research gap identified across medical sensing, imaging, and reconstruction efforts is the predominance of task-specific AI models that are optimized for narrow objectives, such as detection, segmentation, or classification in isolation. 
    While effective for individual tasks, these siloed approaches fail to capture the interconnected nature of clinical decision-making, where diagnosis, risk assessment, treatment planning, and outcome prediction are inherently coupled. 
    The lack of clinical insights-enabled collaborative learning limits model robustness and generalization, as shared information across related tasks remains underutilized. 
    As a result, current systems often provide incomplete or fragmented outputs that require additional manual integration by clinicians, reducing efficiency and hindering adoption in real-world clinical workflows.

    \item \textbf{Limited exploration of non-classical computing paradigms for medical imaging.}
    One research gap highlighted in the workshop is the limited exploration and integration of non-classical computing paradigms, such as quantum computing, in medical sensing, imaging, and reconstruction. 
    Current medical imaging pipelines rely almost exclusively on classical computing architectures, which face increasing challenges in handling the scale and complexity of modern imaging data and reconstruction algorithms. 
    While quantum computing offers theoretical advantages for solving certain classes of optimization, linear algebra, and sampling problems, its potential impact on medical imaging remains largely unexplored. Key gaps include the lack of problem formulations that map clinically relevant imaging tasks to quantum algorithms, insufficient understanding of where quantum advantages may realistically emerge over classical methods, and limited integration between quantum algorithm development and practical medical imaging workflows.
\end{itemize}

\subsection{Future Directions and Recommendations to NSF}

Strategic research directions are needed to enable scalable, trustworthy, and clinically impactful next-generation medical imaging systems. Key priorities include the development of shared hybrid virtual-physical imaging ecosystems, modular and hardware-aware system architectures, and clinically informed collaborative learning frameworks that better reflect real-world medical practice. Together, these directions call for a coordinated rethinking of imaging infrastructure, system design, and computational paradigms, including exploratory advances in quantum computing, to accelerate safe translation from research to clinical deployment.

\begin{itemize}

    
    \item \textbf{Hybrid virtual-physical imaging ecosystems for evaluation and deployment}
    A critical future need identified in the workshop is the development of shared hybrid virtual-physical imaging ecosystems that can serve as community research infrastructure rather than isolated, proprietary systems. 
    While early efforts demonstrate the potential of digital twins for system design and evaluation, there is currently no standardized, open framework for virtual-physical co-evaluation of imaging protocols, reconstruction methods, and clinical workflows. This limits reproducibility, cross-site validation, and regulatory engagement.
    NSF can play a unique role by supporting the creation of open, interoperable digital twins of imaging systems, patients, and clinical environments, 
    including the development of a pioneer population of healthcare avatars that represent diverse anatomies, pathologies, and care contexts,
    along with benchmarks, metrics, and datasets that enable rigorous virtual testing prior to and alongside physical deployment. 
    Such infrastructure would support large-scale validation, safer regulatory science, and broader data sharing, while lowering the barrier for academic and small-industry participation. Investing in shared hybrid virtual-physical ecosystems would enable a step change in how medical imaging technologies are designed, evaluated, and translated to practice.

    \item \textbf{Support modular, flexible, and scalable system architectures for medical imaging.}
    NSF should encourage research that moves beyond tightly coupled, fixed-function medical imaging systems toward architectures built from modular components that can be recomposed as performance requirements, cost constraints, and available hardware evolve,
    enabling reductions in system cost, faster iteration and deployment, and broader availability across care settings. 
    Critically, such architectures should be designed with explicit awareness of memory hierarchies, data movement, and storage constraints on modern accelerators, rather than assuming unlimited on-device resources. 
    As part of this effort, NSF could support the implementation of imaging scanners that explore flexible, software-defined control system architectures, reusable system components with abstracted compute resources, advanced sensor fusion and situational awareness within the exam room, and automatic yet interactive protocol selection informed by available patient information.
    Such architectures should explicitly support heterogeneity in sensing, data movement, and computation,
    including the ability to efficiently utilize GPUs, FPGAs, TPUs, and other specialized accelerators while mitigating memory and storage bottlenecks through hardware-aware scheduling and resource management,
    enabling system functionality to be scaled or reallocated across resources without requiring full redesigns. 
    Priorities include reducing architectural rigidity that limits reuse across product tiers, improving adaptability to hardware obsolescence and supply-chain variability, and enabling efficient real-time processing under diverse deployment conditions. 
    Investment in modular and scalable system architectures would lower development costs, accelerate innovation, improve the long-term sustainability and deployability of medical imaging platforms across clinical and research settings.

    \item \textbf{Enable clinical insights-enabled collaborative and multi-task learning frameworks.}
    NSF should support research that advances clinical insights-enabled collaborative learning, where AI systems are designed to learn jointly from multiple clinically relevant tasks rather than optimizing single objectives in isolation. 
    This includes multi-task and multi-objective learning frameworks that can integrate detection, classification, diagnosis, treatment planning, and outcome assessment within unified models. 
    By encouraging approaches that explicitly model task interdependencies and shared clinical context, NSF can help enable AI systems that generalize more robustly across patients, institutions, and clinical scenarios. 
    Investment in this direction would support more comprehensive, clinically meaningful decision support systems that better reflect real-world medical practice and improve the reliability and utility of AI-assisted care.

    \item \textbf{Support exploratory research on quantum computing for medical imaging and reconstruction.}
    NSF should support exploratory and foundational research investigating the role of quantum computing in medical sensing, imaging, and reconstruction. 
    This includes developing quantum and hybrid quantum-classical algorithms for computationally intensive tasks such as image reconstruction, optimization, and large-scale data analysis, as well as identifying imaging problems where quantum approaches may offer meaningful advantages. 
    NSF investment could enable interdisciplinary collaborations among quantum computing researchers, imaging scientists, and clinicians to align algorithmic advances with real clinical needs. Supporting early-stage benchmarks, simulations, and proof-of-concept studies would help clarify the feasibility, limitations, and long-term potential of quantum computing as a complementary computational paradigm for next-generation medical imaging systems.
    
\end{itemize}

\section{General Future Directions and Recommendations to NSF}
~\label{sec:general}

In addition to the recommendations articulated within each research theme, the following priorities synthesize cross-cutting insights across the four research themes to define a cohesive framework for NSF investment. Collectively, they emphasize a shift away from isolated technological advances toward integrated, resilient, socio-technical medical systems, which are designed to bridge the persistent gap between academic innovation and the realities of clinical deployment and care delivery.

\begin{itemize}

\item \textbf{Prioritize human-centered resilience over isolated algorithmic accuracy.} Workshop participants emphasized that, in safety-critical medical contexts, system reliability under stress is a more vital engineering objective than peak accuracy in controlled settings. NSF should support research into 
failure-aware
(RT3) and graceful degradation (RT2) models that maintain clinical utility when hardware faults occur, sensors are displaced, or connectivity is lost. This includes advancing just-enough interaction (RT1, RT4) frameworks that prioritize human agency and expert-in-the-loop accountability, ensuring robustness in real-world deployments.

\item \textbf{Enable de-risking pathways to bridge the translational ``valley of death.''}
To accelerate the transition from laboratory prototypes to commercially viable medical products, NSF should fund pre-clinical innovation pathways that incorporate manufacturability, regulatory science, and economic feasibility into early-stage research and development. Leveraging directorates such as Technology, Innovation and Partnerships (TIP), programs should incentivize interdisciplinary teams to address hardware security and IRB considerations early in the design process. This system-level focus ensures that technical merit is coupled with a realistic roadmap for clinical and market entry.

\item \textbf{Cultivate open ecosystems for research-ready longitudinal and social data.} Current clinical datasets, often optimized for documentation and billing, lack the granularity required to engineer robust AI systems. NSF should support the development of standardized, shared datasets that capture real-world failure cases, longitudinal patient behaviors, and social determinants of health. By expanding access to high-quality, multimodal data, similar in spirit to UK Biobank and AI-READI, NSF can provide the community with the ground truth needed to train models that are representative of diverse populations and complex clinical workflows.

\item \textbf{Catalyze closed-loop socio-biological systems for holistic patient care.} Future research should move beyond siloed physiological monitoring to integrate mental, emotional, and social health into the treatment loop. NSF should encourage the development of biopsychosocial controllers, AI systems that unify living pharmacies (RT2) with social health sensing (RT3). Such systems could adapt 
therapeutic systems’ biochemical dosage based on objective digital biomarkers of social isolation or cognitive load, treating the patient as a dynamic, integrated biological and social entity.

\item \textbf{Develop scale-invariant and cross-modality foundation models for medicine.} As medical AI models grow in complexity, NSF should support the 
development of multi-modality, multi-task, generalizable foundation models that learn shared representations across medical modalities and scales, spanning micro-scale neural sensing (RT2) to macro-scale 4D imaging (RT4).
These approaches should enable hardware-adaptive deployments, allowing model capacity and computational cost to scale with available resources, from sub-milliwatt implantable systems-on-chip to high-end radiology workstations. 
Such cross-modality reasoning could support predictive systems that connect subtle home-based behavioral signals (RT3) with emerging surgical or cardiac risks (RT1).

\item \textbf{Promote privacy-by-design for the hybrid medical edge.} The proliferation of medical Internet-of-Things technologies across the care continuum necessitates standardized architectures for secure data sharing. NSF should prioritize privacy-by-design learning and deployment frameworks that enable transmit-information-not-data (RT2) strategies. By supporting secure on-site processing and formal guarantees against information leakage, NSF can enable hybrid edge ecosystems in which implants, home sensors, and hospital scanners share insights without exposing raw patient data.

\item \textbf{
Establish virtual-physical ecosystems and trustworthy generative AI.}
As generative models are increasingly used to augment limited datasets and synthesize rare, safety-critical scenarios, NSF should support rigorous methods for validating clinical realism and downstream impact. This includes developing standardized protocols and physics-informed benchmarks to ensure that generated images and videos (RT4), as well as simulated home environments (RT3), provide high-fidelity training signals that translate to real-world clinical performance.
NSF should further enable a virtual-physical healthcare ecosystem, including diverse healthcare avatars, representing diverse anatomies, pathologies, and care contexts, and shared benchmarks, metrics, and datasets, to support rigorous virtual testing and validation prior to and alongside real-world deployment.

\item \textbf{Strengthen shared compute infrastructure and open foundation model ecosystems.} To address the widening resource gap between academia and industry, NSF should play a central role in supporting shared compute infrastructure and community-scale testbeds. This includes providing cloud credits and access to specialized hardware required for training and deploying large-scale medical foundation models. By promoting openly licensed and transparent model ecosystems, NSF can enable reproducible academic innovation without reliance on proprietary, closed-form systems.

\item \textbf{Support modular, software-defined, and scalable system architectures.} NSF should promote research into modular medical architectures that can be recomposed as performance requirements and hardware resources evolve. Advancing software-defined scanners (RT4) and adaptive control systems (RT1) will reduce development costs and mitigate hardware obsolescence. These architectures must be hardware-aware, efficiently managing memory hierarchies and data movement across heterogeneous accelerators (e.g., GPUs, FPGAs, and TPUs) to ensure high-performance medical AI is deployable across all tiers of healthcare delivery.

\end{itemize}

\section{Conclusion}

The future of healthcare demands a fundamental shift away from isolated software or hardware advances toward a resilient, systems-level philosophy that tightly integrates algorithm-hardware co-design with the intrinsic uncertainty of human health and care delivery. Grounded in the converging requirements of surgical teleoperation, implantable living pharmacies, home-based ICUs, and next-generation medical imaging, this roadmap calls for sustained federal investment in: 
(i) modular, software-defined system architectures that ensure long-term dependability; 
(ii) objective digital biomarkers that translate subjective patient experiences into clinically actionable signals; 
(iii) human-aware AI systems that foster trust across the full care continuum; 
and (iv) virtual-physical validation ecosystems that couple simulation, generative modeling, and representative healthcare avatars with physical deployment to enable safe, scalable, and reproducible translational testing. 
Leveraging the NSF’s unique capacity to de-risk translational research and overcome the lab-to-deployment gap, this approach bridges academic innovation and commercial deployment, transforming today’s laboratory prototypes into tomorrow’s dependable, secure foundations of clinical care.

\IEEEpubidadjcol

\section*{Acknowledgments}

This workshop was supported by the National Science Foundation under Award \#2337454.
We sincerely thank the invited speakers, Christopher Scully, Jonathan S. Maltz, Sharon Xiaolei Huang, Ajmal Zemmar, Danny Chen, Euisik Yoon, George Demiris, Jie Gu, Milan Sonka, Mohammed Islam, Omer Inan, Robert Webster III, Roozbeh Jafari, Sam Kavusi, and Shandong Wu, for sharing their expertise and insights. 
We also thank the panel moderators, Wei Ding, Mohammed Islam, Mingmin Zhao, and Shandong Wu, and the panelists, Aidong Zhang, Sam Kavusi, Robert Webster III, Yiyu Shi, Roozbeh Jafari, Jie Gu, Mohammed Islam, Euisik Yoon, Miroslav Pajic, George Demiris, Omer Inan, Mingmin Zhao, Swarun Kumar, Shandong Wu, Xiaofeng Yang, Howie Choset, Justin Chan, and Sharon Xiaolei Huang, 
and James Weimer, 
for leading thoughtful discussions and contributing valuable perspectives. 
We are grateful to the steering and organizing committee members, 
Insup Lee, Aidong Zhang, Robert Dick, Xiaodong Wu, Majid Sarrafzadeh, Yiyu Shi, 
Peipei Zhou, and Jingtong Hu, for their guidance and service. 
We thank the breakout discussion leads, Jinglong Hu, Jeff Zhang, Xiaofeng Yang, Shandong Wu, Jie Gu, Zongxing Xie, Mingmin Zhao, and Kuk Jin Jang, for facilitating the interactive breakout session discussion, and Zheng Dong, Zhuoping Yang, and Weisong Shi, for their constructive contributions to the preparation of this report. 
We gratefully acknowledge the support and guidance of the NSF program directors, X. Sharon Hu, Goli Yamini, and Sankar Basu.
Finally, we thank the Pitt Engineering staff, Melissa Penkrot, Justin Misera, John Spadone, and Jessica Dawson, for their valuable assistance with workshop logistics, grant administration, and travel reimbursement, and Brown and Pitt student volunteers, Zhuoping Yang, Wei Zhang, Jingxian Chen, Zhimin Li, Pan Wang, and Yukai Song, who assisted and supported throughout the workshop.



\bibliographystyle{IEEEtran}
\balance
\bibliography{bib/ref_workshop}

\vfill

\end{document}